\documentclass[a4paper,11pt]{article}
\pdfoutput=1

\usepackage{jheppub}
\usepackage[T1]{fontenc}
\usepackage{amssymb,amsmath}
\usepackage{braket,bbm,bm}
\usepackage{mathtools,amssymb,dsfont}
\usepackage[usenames,dvipsnames,svgnames,table]{xcolor}
\usepackage[utf8]{inputenc}
\usepackage{soul, color}
\usepackage{subcaption}
\usepackage{lmodern,enumitem}
\usepackage{footnote}
\usepackage[normalem]{ulem}
\usepackage{glossaries-extra}
\usepackage{slashed}
\usepackage{multirow}
\usepackage{here}
\usepackage{hyperref}
\usepackage{verbatim}
\usepackage[justification=justified,singlelinecheck=false]{caption}
\usepackage{cleveref}
\usepackage{listings}
\usepackage{fancyvrb,placeins}
\usepackage{dsfont}
\usepackage{nicefrac,xfrac}
\usepackage{bbm}

\Crefname{equation}{eq.}{eqs.}
\Crefname{section}{section}{sections}
\Crefname{figure}{figure}{figures}
\Crefname{appendix}{appendix}{appendices}

\newcommand{\vx}{\mathbf{x}}
\newcommand{\vy}{\mathbf{y}}
\newcommand{\vp}{\mathbf{p}}
\newcommand{\vq}{\mathbf{q}}
\newcommand{\vP}{\mathbf{P}}
\newcommand{\vd}{\mathbf{d}}

\newcommand{\Amp}{{\cal A}_{\mathbf{27}}}
\newcommand{\SUF}{SU(3)$_{\rm F}$}
\newcommand{\acp}{\Delta A_{\rm CP}}

\hypersetup{
pdftitle = {$D \to K \pi$ at the SU(3)-flavour-symmetric point},
pdfsubject = {$D \to K \pi$ at the SU(3)-flavour-symmetric point},
pdfkeywords = {lattice QCD, hadron decays, charm decays},
pdfauthor = {Black et al.},
pdfnewwindow = {true},
colorlinks = {true},
linkcolor = {blue},
citecolor = {blue},
filecolor = {blue},
urlcolor = {blue}
}

\title{\boldmath $D \to (K \pi)_{\mathbf{27}}$ at the SU(3)-flavour-symmetric point\\ {\rm\bf I}: Methodology and strong phase determination}

\author[a]{Matthew Black,}
\author[b]{Felix Erben,}
\author[a]{Maxwell T.~Hansen,}
\author[c,a]{Fabian Joswig,}
\author[d]{Nelson Pitanga Lachini,}
\author[a]{Rajnandini Mukherjee,}
\author[e]{Srijit Paul,}
\author[a]{and Antonin Portelli}

\affiliation[a]{School of Physics and Astronomy, University of Edinburgh, Edinburgh EH9 3JZ, UK}

\affiliation[b]{CERN, Theoretical Physics Department, CH-1211 Geneva 23, Switzerland}

\affiliation[c]{DeepL SE, Maarweg 165, 50825 Köln, Germany}

\affiliation[d]{DAMTP, University of Cambridge, Wilberforce Road, Cambridge, CB3 0WA, UK}

\affiliation[e]{Department of Physics, University of Cyprus, Aglantzia 2109, Nicosia, Cyprus}

\emailAdd{matthew.black@ed.ac.uk}
\emailAdd{felix.erben@cern.ch}
\emailAdd{maxwell.hansen@ed.ac.uk}
\emailAdd{fabian.joswig@deepl.com}
\emailAdd{np612@cam.ac.uk}
\emailAdd{r.mukherjee@ed.ac.uk}
\emailAdd{srijitpaul@gmail.com}
\emailAdd{antonin.portelli@ed.ac.uk}

\abstract{We present part one of an SU(3)-flavour-symmetric lattice QCD calculation of the amplitude for a $D$-meson decaying to a $K\pi$ final state in the 27-dimensional irreducible representation of the flavour symmetry group, denoted $(K\pi)_{\mathbf{27}}$.
The Wilson--clover gauge ensembles used in this work, generated by the OpenLat collaboration, are tuned such that $M_\pi = M_K \approx 410\,\mathrm{MeV}$.
Using the distillation framework, we construct a matrix of Euclidean correlation functions from pairs of single-hadron operators projected to definite spatial momentum.
Solving a generalised eigenvalue problem yields the finite-volume energy spectrum that is used to determine the scattering phase shift from threshold up to $4 M_\pi \approx 1640 \,\mathrm{MeV}$, which sits below but plausibly within reach of $M_D^{\rm SU(3)} \simeq 1900\,\mathrm{MeV}$.
The calculation is performed across three lattice spacings, and we apply two strategies in which the continuum limit is taken at different stages of the computation: (i) on the extracted scattering parameters and (ii) on the finite-volume energies at fixed physical volume before extracting the scattering parameters.
We find consistent results across these methods for the scattering phase shift as a function of the centre-of-mass energy, $\delta_{\mathbf{27}}(E_{\sf cm})$. Taking a scattering-length-only parametrisation, we infer a value for the strong phase of the weak decay, $\delta_{\mathbf{27}}(M_D^{\rm SU(3)})=-38.4(2.4)^\circ$.
We further describe the methodology for using the same operator basis to compute three-point correlation functions to extract $\langle (K\pi)_{\mathbf{27}}| H_W| D\rangle$, for the tree-level effective weak Hamiltonian $H_W$, and for relating such finite-volume matrix elements to the full decay amplitude. 
The complete analysis leading to the latter will be presented in a forthcoming manuscript.
}

\preprint{CERN-TH-2026-204}

\begin{document}
\maketitle
\flushbottom

\section{Introduction}
\label{sec:intro}

Nonleptonic $D$-meson decay amplitudes are an important class of observables for testing the flavour structure of the Standard Model (SM) and for searching for new sources of charge-parity (CP) violation.
Charm is the only up-type quark to form hadrons that decay weakly to other hadrons, enabling precision studies of such transitions. The charm sector thus provides a natural complement to the kaon and bottom systems, where CP violation has long been established.
In the SM, direct CP violation in singly Cabibbo-suppressed (SCS) charm decays is expected to be strongly suppressed by the CKM hierarchy and by the interplay of short-distance weak phases with long-distance QCD dynamics~\cite{Lenz:2020awd}. As a result, interpreting measurements to high precision requires quantitative control of nonperturbative hadronic amplitudes~\cite{Grossman:2006jg,Brod:2011re,Grossman:2012eb,Franco:2012ck}.
General reviews of hadronic $D$ and $D_s$ decays, together with data-driven analyses of two-body $D$ decays based on topological amplitudes, flavour breaking, final-state interactions, and factorisation-assisted approaches, can be found in refs.~\cite{Ryd:2009uf,Chiang:2002mr,Cheng:2010ry,Qin:2021tve}.

A particularly compelling measure of CP violation in hadronic charm decays is the time-integrated CP asymmetry for a final state $f$, defined as
\begin{equation}
    A_{CP}(f) = \frac{\Gamma(D^0 \to f) - \Gamma(\bar D^0 \to f)}{\Gamma(D^0 \to f) + \Gamma(\bar D^0 \to f)}.
\end{equation}
Forming differences between asymmetries for different final states significantly reduces experimental systematic uncertainties, motivating, for example, the observable
\begin{equation}
\acp \equiv A_{CP}(  K^+K^-) - A_{CP}( \pi^+\pi^-).
\end{equation}
This quantity is expected to be numerically suppressed, primarily because both decay channels are singly Cabibbo-suppressed. Naive perturbative methods~\cite{Grossman:2006jg,Bigi:2011re,Lenz:2013pwa,Grossman:2019xcj} lead to the upper bound
\begin{equation}
    |\acp|^{\rm naive} \lesssim 2.6\times10^{-4} \,.
\end{equation}
To better control the non-perturbative effects of these amplitudes, light-cone sum rules (LCSR)~\cite{Balitsky:1989ry} have also been used for these decays~\cite{Khodji:2017zdu,Chala:2019fdb,Lenz:2023rlq,Li:2012cfa}.
These methods 
yield similarly suppressed results
with the latest determination of ref.~\cite{Lenz:2023rlq} finding
\begin{equation}
    |\acp|^{\rm LCSR} \leq 2.4\times10^{-4}.
\end{equation}

Despite these expectations, LHCb first reported evidence for a nonzero $\acp$ in 2011~\cite{LHCb:2011osy}.
Using the full Run~1 and Run~2 data sets of the Large Hadron Collider at CERN, the collaboration confirmed the discovery of CP violation in the charm sector in 2019~\cite{LHCb:2019hro}, finding
\begin{equation}
    \acp = (-15.4 \pm 2.9)\times 10^{-4},
\end{equation}
a result whose magnitude significantly exceeds the aforementioned theoretical bounds.
It is well accepted, however, that these predictions are incomplete and greater theoretical understanding of hadronic charm decays is needed before one can identify the origin of $\acp$.
Using U-spin and \SUF\ symmetry arguments, refs.~\cite{Grossman:2019xcj,Cheng:2019ggx} both conclude the value of $\acp$ can be explained within the SM via some mild \SUF-breaking effects generated either perturbatively or non-perturbatively by e.g.\ final-state scattering effects or even resonance enhancements~\cite{Schacht:2021jaz,Bediaga:2022sxw}.

Inspired by this situation, in this and a subsequent manuscript (to appear), we present a lattice QCD calculation of a simplified observable that represents a step towards the full physical prediction: the $D\to(K\pi)_{\mathbf{27}}$ decay amplitude in \SUF\ QCD. In the present manuscript, we detail our calculation of the $K\pi$ scattering amplitude in the 27-plet channel, which gives the strong phase of the weak decay amplitude and is a prerequisite of the full calculation for other technical reasons related to the lattice method.
To determine the hadronic scattering amplitude, we calculate the finite-volume $K\pi$ spectrum using distillation~\cite{HadSpec:2009krc} and a variational analysis based on the generalised eigenvalue problem (GEVP)~\cite{Luscher:1990ck,Blossier:2009kd} on three \SUF\ lattice gauge ensembles at three different lattice spacings generated by the OpenLat collaboration~\cite{Francis:2022hyr,Cuteri:2022oms}.
We then apply L\"uscher's quantisation condition~\cite{Luscher:1986pf,Luscher:1990ux} to extract the elastic phase shift and its continuum limit.

Looking towards the full calculation, the general established method is to use the same operator basis to construct projected three-point correlation functions containing the effective weak Hamiltonian and extract the corresponding finite-volume matrix elements. These are then related to the infinite-volume decay amplitudes through the Lellouch--L\"uscher formalism~\cite{Lellouch:2000pv}, using the scattering information determined from the finite-volume spectrum.
In this manuscript, in addition to the complete analysis of the finite-volume energies and the corresponding hadronic scattering amplitude, we describe the method for evaluating these three-point functions within the distillation framework, including an appraisal of the associated computational and storage costs. We also evaluate the required Lellouch--L\"uscher factors. Our implementation of the three-point-function analysis, including the complete determination of the weak matrix elements and resulting decay amplitude, will be presented in a part two.

Lattice calculations of hadronic decays to two-particle final states have previously been carried out most extensively for $K\to\pi\pi$, including continuum-limit results for the $I=2$ amplitude and a determination of the $I=0$ amplitude at physical kinematics by the RBC/UKQCD collaboration~\cite{Blum:2015ywa,RBC:2020kdj}. More recently, the first lattice calculation of the semileptonic $B\to\pi\pi\ell\bar\nu$ amplitude has determined $B\to\rho\ell\bar\nu$ form factors while treating the $\rho$ as a resonance, by applying the finite-volume formalism and analytically continuing the amplitude to the $\rho$ pole~\cite{Leskovec:2025gsw}.
Similar work is ongoing to compute the flavour-changing neutral-current form factors for the decay $B\to K^*\ell\bar\nu$~\cite{Erben:2026ylp}. Related applications of the finite-volume formalism include lattice determinations of the $\pi\gamma\to\pi\pi$ and $K\gamma\to K\pi$ transition amplitudes, giving access to the resonant radiative decays $\rho\to\pi\gamma$ and $K^*\to K\gamma$~\cite{Briceno:2016kkp,Alexandrou:2018jbt,Radhakrishnan:2022ubg}.

Before closing the introduction, we stress here that our calculation of the $D \to (K \pi)_{\textbf{27}}$ decay amplitude is a first step on a long-term journey to ultimately give reliable predictions for such amplitudes with fully-controlled physical-mass calculations. In the context considered here, in which a finite-volume formalism is used, the lattice calculation of a decay amplitude involving a multi-hadron final state proceeds through the following general workflow:
\begin{enumerate}
    \item Select the decay channel and kinematic region of interest, and identify all multi-hadron channels that are open over the target range of centre-of-mass energies.
    \item For each open channel, construct multi-hadron interpolating operators whose constituent hadrons are projected to definite spatial momenta. The operators are organised into irreducible representations of the relevant internal symmetries and of the finite-volume little group. At each physical volume, the basis is chosen using the non-interacting finite-volume levels that populate the target energy range and is supplemented by single-hadron-like operators in channels where resonances are expected.
    \item Using this operator basis, compute a matrix of two-point correlation functions, typically employing distillation or another efficient all-to-all method~\cite{HadSpec:2009krc}.
    \item Solve a generalised eigenvalue problem to reliably determine the finite-volume energy spectrum~\cite{Luscher:1990ck,Blossier:2009kd}, ideally for multiple finite-volume irreducible representations, total momenta and spatial volumes.
    \item Analyse the spectrum using a finite-volume scattering formalism that accommodates all open channels, thereby determining the coupled hadronic scattering amplitudes relevant to the system.
    \item Use the optimised operators obtained from the variational analysis to construct three-point correlation functions involving the incoming parent hadron (here the $D$ meson) and the transition operator. 
    \item Fit the three-point correlation functions to extract the corresponding finite-volume transition matrix elements.
    \item Apply the Lellouch--L\"uscher formalism, or its appropriate generalisation, using the hadronic scattering amplitudes as input, to map the finite-volume matrix elements to the physical decay amplitudes~\cite{Lellouch:2000pv}.
\end{enumerate}

This overview also serves to establish the many challenges of such a calculation, especially for physical pion masses. For example, with regards to items 2-4, the set of open channels for vacuum quantum number multi-hadron channels that can  propagate for energies corresponding to the $D$-meson mass includes at least the following: $\pi\pi, $ $
4\pi,  $ $
6\pi,  $ $
K\bar K,  $ $
\eta\eta,  $ $
4\pi\eta,  $ $
8\pi, $ $
K\bar K\,\pi \pi,  $ $
\eta\eta\,\pi \pi,  $ $
6\pi\eta,  $ $
10\pi,  $ $
K\bar K\,3\pi,  $ $
K\bar K\,4\pi,  $ $
\eta\eta\,4\pi,  $ $
8\pi\eta,  $ $
12\pi,$ $ 
K\bar K\, \pi\eta, $ and $
K\bar K\,5\pi$. This also requires a highly sophisticated finite-volume formalism in items 5 and 8, one which, in fact, has not yet been fully developed given the requirement to include channels with more than three hadrons.
Work is ongoing to develop a perturbative $N$-body scattering framework as a first step towards categorising the effects of these higher-multiplicity multi-hadron channels~\cite{Mukherjee:2026fourpion}.
Additional challenges include reliably controlling discretisation effects, especially given the inclusion of the charm quark, and appropriately renormalising the operator, in particular as cases can arise with power-divergent mixing.
In light of these extensive computations we have elected to approach the problem in a series of controlled steps, beginning with a complete determination of the \SUF\ system.

The remainder of this manuscript is organised as follows: \cref{sec:group_theory} reviews the relevant \SUF\ group theory, \cref{sec:kpi_scattering} presents the $K\pi$ scattering analysis, and \cref{sec:dkpi_decay} describes the decay calculation, before we conclude in \cref{sec:conclusions}.

\section{\SUF\ group theory}
\label{sec:group_theory}

As discussed above, flavour symmetry is significantly extended in the case that $m_u = m_d = m_s$ (equivalently $M_\pi = M_K = M_\eta$), leading to an exact \SUF\ symmetry.
To meaningfully design calculations in this limit, and to interpret the results, it is essential to understand how the various hadronic states and weak operators transform under this enlarged symmetry group.

We begin by grouping the three lightest quarks into a \SUF\ triplet,
\begin{equation} \label{eq:suftriplet}
    \boldsymbol{\ell}^T  = (u,d,s),
\end{equation}
while all other quarks, importantly here including the charm, are singlets.
More precisely we say that $\boldsymbol{\ell}$ transforms in the fundamental representation $\mathbf{3}$ of \SUF\ defined by $ \boldsymbol{\ell} \to U \boldsymbol{\ell}$ where $U$ is a $3\times 3$ special unitary matrix.

We now consider the transformation properties inherited by the various hadronic states and weak operators of interest.

Beginning with the incoming $D$ meson, the six charmed mesons of relevance group into two triplets
\begin{gather}
    \vert \boldsymbol D \rangle =
        \left( \vert D^0 \rangle = \vert c\bar{u} \rangle, \ \ \vert D^+ \rangle = \vert c\bar{d} \rangle, \ \ \vert D_s^+ \rangle = \vert c\bar{s} \rangle \right)  \in \ \overline{\mathbf{3}} \,, \\
    \vert \overline{\boldsymbol D} \rangle =
        \left( \vert \bar{D}^0 \rangle = \vert \bar{c}u \rangle, \ \ \vert D^- \rangle = \vert \bar{c}d \rangle, \ \ \vert D_s^- \rangle = \vert \bar{c}s \rangle \right)  \in \ \mathbf{3} \,,
\end{gather}
where $\in \overline{\mathbf{3}}$ and $\in \mathbf{3}$ indicate states that respectively transform in the anti-fundamental and fundamental representations of \SUF.
The operators that create these states from the vacuum are then
\begin{gather}
    \boldsymbol {\mathcal O}_{D} = \left( \bar{c} \gamma_5 u, \ \bar{c} \gamma_5 d, \ \bar{c} \gamma_5 s \right)  \in \mathbf{3} \,,\\
    \overline{\boldsymbol {\mathcal O}}_{{D}} = \left( \bar{u} \gamma_5 c, \ \bar{d} \gamma_5 c, \ \bar{s} \gamma_5 c \right)  \in \overline{\mathbf{3}} \,.
\end{gather}
As usual, the operators transform in the conjugate representation to the incoming states they create.

Next we consider the $K \pi$ final state.
Both the kaon and pion are made from a light quark and antiquark, and thus both transform in the $\mathbf{3} \otimes \overline{\mathbf{3}}$ representation.
This decomposes into irreducible representations (irreps) as
\begin{equation}
    \mathbf{3} \otimes \overline{\mathbf{3}} = \mathbf{1} \oplus \mathbf{8} \,,
\end{equation}
and the pseudoscalar mesons belong to the octet representation, also known as the adjoint representation.
It is convenient and standard to represent the octet as a $3\times 3$ traceless matrix $\Phi$ that transforms as $\Phi \to U \Phi U^\dagger$.
In terms of the individual meson fields, the latter is given by~\cite{deSwart:1963pdg}
\begin{equation}
    \Phi = \left(
            \begin{array}{ccc}
            \frac{1}{\sqrt{2}}\pi^0 + \frac{1}{\sqrt{6}}\eta_8 & \pi^+ & K^+ \\
            \pi^- & -\frac{1}{\sqrt{2}}\pi^0 + \frac{1}{\sqrt{6}}\eta_8 & K^0 \\
            \bar{K}^- & \bar{K}^0 & -\sqrt{\frac{2}{3}}\,\eta_8
            \end{array}
        \right) \in \mathbf{8}.
\end{equation}

The $K\pi$ final state therefore belongs to the decomposition of the octet tensor product with itself:
\begin{equation}
\mathbf{8}\otimes \mathbf{8}
= \mathbf{1}\oplus \mathbf{8}_S \oplus \mathbf{8}_A \oplus \mathbf{10}\oplus \overline{\mathbf{10}}\oplus \mathbf{27} \,.
\end{equation}
In the context of a hadronic $D$ decay, we are only interested in the two-hadron states that can overlap the product of the incoming $D$ meson and the weak Hamiltonian.
Both of these have zero angular momentum and thus the same must be true for the final state.
This, together with the exchange symmetry of bosons, implies that only the states symmetric under the exchange of the two octets can contribute, i.e.
\begin{equation} \label{eq:8x8irreps}
    ( \mathbf{8}\otimes \mathbf{8} )_S = \mathbf{1}\oplus \mathbf{8}_S \oplus \mathbf{27} \,.
\end{equation}
So, without having yet considered the weak Hamiltonian, we see that six categories of decays emerge:
\begin{gather}
    \boldsymbol{D} \to (\Phi \Phi)_{\mathbf{1}}, \quad
    \boldsymbol{D} \to (\Phi \Phi)_{\mathbf{8}_S}, \quad
    \boldsymbol{D} \to (\Phi \Phi)_{\mathbf{27}}, \\
    \overline{\boldsymbol{D}} \to (\Phi \Phi)_{\mathbf{1}}, \quad
    \overline{\boldsymbol{D}} \to (\Phi \Phi)_{\mathbf{8}_S}, \quad
    \overline{\boldsymbol{D}} \to (\Phi \Phi)_{\mathbf{27}}.
\end{gather}
Particular $K\pi$ channels can be located inside this decomposition using isospin $I$ and hypercharge $Y$.
Kaons have $I=1/2$ and $Y=\pm1$ while pions have $I=1$ and $Y=0$.
Thus a $K\pi$ or $\bar{K}\pi$ state has
\begin{equation}
    Y = \pm 1,\qquad I = \frac12\otimes1 = \frac12\oplus\frac32.
\end{equation}
In \cref{eq:8x8irreps}, the $\mathbf{1}$ irrep contains only $Y=0,\,I=0$ and therefore contains no $K\pi$ state.
On the other hand, the $\mathbf{8}_S$ contains $Y=\pm 1$ hypercharges and $I=1/2$ and the $\mathbf{27}$ contains the largest set of $K\pi$ channels with $Y=\pm1$ and $I=1/2,3/2$.
Therefore to project to $K\pi$ states with definite isospin, there are two possibilities.
A $K\pi$ state with $I=1/2$ gets contributions from both the octet and 27-plet,
\begin{equation} \label{eq:KpiOctet}
    |K\pi;\,I=\tfrac12\rangle = c_8\,|K\pi;\,I=\tfrac12\rangle_{\mathbf{8}} + c_{27}\,|K\pi;\,I=\tfrac12\rangle_{\mathbf{27}},
\end{equation}
while a state with maximal isospin $I=3/2$ can be extracted exclusively from the 27-plet:
\begin{equation} \label{eq:Kpi27plet}
    |K\pi;\,I=\tfrac32\rangle = |K\pi;\,I=\tfrac32\rangle_{\mathbf{27}},
\end{equation}
up to normalisation and phase conventions.
For example, focusing on the $Y=-1$ sector and thus $\bar{K}\pi$ states, the maximal-isospin multiplet with $I=3/2$ contains four states corresponding to different values of $I_3$.
These are
\begin{align}
    |\bar{K}\pi;\,I=\tfrac32,\,I_3=+\tfrac32\rangle &= |\bar{K}^0\pi^+\rangle \,, \label{eq:KpiUsed} \\
    |\bar{K}\pi;\,I=\tfrac32,\,I_3=+\tfrac12\rangle &= \sqrt{\frac13}|K^-\pi^+\rangle + \sqrt{\frac23}|\bar{K}^0\pi^0\rangle \,, \\
    |\bar{K}\pi;\,I=\tfrac32,\,I_3=-\tfrac12\rangle &= \sqrt{\frac13}|\bar{K}^0\pi^-\rangle + \sqrt{\frac23}|K^-\pi^0\rangle \,, \\
    |\bar{K}\pi;\,I=\tfrac32,\,I_3=-\tfrac32\rangle &= |K^-\pi^-\rangle \,.
\end{align}
Only the extremal states with $I_3=\pm\frac32$ correspond to single flavour-specific final states. Charge conservation, however, forbids $K^-\pi^-$ in $D$-meson decays, leaving $\bar{K}^0\pi^+$ as the unique directly accessible pure $I_3 = \frac 32$ state. That is, in an \SUF\ symmetric world, the decay amplitude for $D^+ \to \bar K^0 \pi^+$ is equal to the definite final-state irrep amplitude $D \to (K \pi)_{\mathbf{27}}$.

Now turning to the final missing component of the full decay, and focusing on the generic $\Delta C = 1$ weak Hamiltonian (without any assumption of Cabibbo sector), we write
\begin{equation}
    \mathcal{H}^{\Delta C = 1}_{\text{eff}} \supset \sum_{i,\,j,\,k} C^{ij}{}_{k}  {\cal O}^{ij}{}_{k} \,,
\end{equation}
where the sum is over the flavour indices $i,\,j,\,k$ and we have defined the current-current operators
\begin{equation} \label{eq:OpTensor}
    {\cal O}^{ij}{}_{k} = (\bar{\boldsymbol \ell}^{i}\,\Gamma_1\,c)(\bar{\boldsymbol \ell}^{j}\,\Gamma_2\,{\boldsymbol \ell}_{k}) \,,
\end{equation}
where $\Gamma_{1,2}$ are left-handed Dirac matrices.
The Wilson coefficients $C^{ij}{}_{k}$ are determined by the short-distance physics and are not constrained by the flavour symmetry, but the transformations of the operators ${\cal O}^{ij}{}_{k}$ are well-defined under \SUF.
These can be decomposed into irreps by decomposing the two bilinears separately and combining the results; see e.g.~\cite{He:2018joe,Petrov:2024ujw}.
The first bilinear transforms as $\overline{\mathbf{3}}$, while the second transforms as $\overline{\mathbf{3}} \otimes {\mathbf{3}} =  \mathbf{1} \oplus \mathbf{8}$.
One finds
\begin{equation}
    {\cal O}^{ij}{}_{k} \in \overline{\mathbf{3}} \otimes \overline{\mathbf{3}} \otimes {\mathbf{3}} = \overline{\mathbf{3}}_{\mathbf 1} \oplus \overline{\mathbf{3}}_{\mathbf 8} \oplus \mathbf{6} \oplus \overline{\mathbf{15}} \,. \label{eq:HamIrreps}
\end{equation}
The most obviously troublesome irrep is the $\overline{\mathbf{3}}_{\mathbf{1}}$ as this arises from the singlet decomposition of the second bilinear in the weak Hamiltonian and naturally contains the flavour-singlet penguin contractions $\sum_q \bar{q}\Gamma q$.
Similarly, the $\overline{\mathbf{3}}_{\mathbf{8}}$ is still a trace component of the full tensor ${\cal O}^{ij}{}_{k}$.
Via Fierz rearrangement of the anti-fundamental light quarks, it is also associated with a singlet contraction.
This distinction is important for lattice calculations as singlet contractions in the four-quark operator lead to mixing with lower-dimensional operators and hence to power-divergent subtractions in the continuum limit.

The remaining two irreps of the weak Hamiltonian, $\mathbf{6}$ and $\overline{\mathbf{15}}$, are respectively the antisymmetric and symmetric traceless parts of the operator tensor ${\cal O}^{ij}{}_k$ in~\cref{eq:HamIrreps}.
These are both `clean' renormalisation targets compared to the $\overline{\mathbf{3}}_{\mathbf{1},\mathbf{8}}$ irreps since, by definition of being traceless, they cannot contain operators with self-contractions and therefore do not mix with lower-dimensional operators.

Finally, we can connect all the elements of our decay process and classify a set of {\it reduced matrix elements} contributing to each final-state irrep from each weak Hamiltonian irrep.
Collecting all non-zero contributions, we reach the following seven decay amplitudes:
\begin{gather}
    \label{eq:DtoPhiPhi1}
    \boldsymbol{D} \underset{\overline{\mathbf{3}}_{\mathbf{1}}}{\longrightarrow} (\Phi \Phi)_{\mathbf{1}}, \quad
    \boldsymbol{D} \underset{\overline{\mathbf{3}}_{\mathbf{8}}}{\longrightarrow} (\Phi \Phi)_{\mathbf{1}}, \\
    \label{eq:DtoPhiPhi8}
    \boldsymbol{D} \underset{\overline{\mathbf{3}}_{\mathbf{1}}}{\longrightarrow} (\Phi \Phi)_{\mathbf{8}_S},\quad
    \boldsymbol{D} \underset{\overline{\mathbf{3}}_{\mathbf{8}}}{\longrightarrow} (\Phi \Phi)_{\mathbf{8}_S},\quad
    \boldsymbol{D} \underset{{\mathbf{6}}}{\longrightarrow} (\Phi \Phi)_{\mathbf{8}_S}, \quad
    \boldsymbol{D} \underset{\overline{\mathbf{15}}}{\longrightarrow} (\Phi \Phi)_{\mathbf{8}_S} \\
    \label{eq:DtoPhiPhi27}
    \boldsymbol{D} \underset{\overline{\mathbf{15}}}{\longrightarrow} (\Phi \Phi)_{\mathbf{27}}\,.
\end{gather}
The Wigner-Eckart theorem states that decay amplitudes in fixed final-state irreps are linear combinations of the reduced matrix elements in~\cref{eq:DtoPhiPhi1,eq:DtoPhiPhi8,eq:DtoPhiPhi27} weighted by SU(3) Clebsch-Gordan coefficients.
The singlet and octet final-state irreps are more complex since they receive contributions from several Hamiltonian irreps, as well as resonance-like contractions.
In contrast, the 27-plet final state is reached only through the $\overline{\mathbf{15}}$ irrep of the Hamiltonian.
Combining this with the fact that the maximal-isospin $K\pi$ states belong exclusively to the 27-plet with no resonance contamination, the decay process in~\cref{eq:DtoPhiPhi27} is an appealing first reduced matrix element to target since it delivers a decay amplitude without requiring a mixture of multiple irreps or introducing challenging renormalisation structures.
As a first step towards a full categorisation of hadronic $D$ decays in the \SUF\ limit, we will therefore focus on computing $\Amp = {\cal A}(\boldsymbol{D} \underset{\overline{\mathbf{15}}}{\longrightarrow} (\Phi \Phi)_{\mathbf{27}})$.

So far, we have discussed the $\Delta C=1$ weak Hamiltonian in general terms.
However, it is most commonly split into subsectors based on the level of CKM-suppression present: Cabibbo Favoured (CF), Singly-Cabibbo Suppressed (SCS), and Doubly-Cabibbo Suppressed (DCS).
For example, SCS transitions involve $c\to d\bar{d}u$ and $c\to s\bar{s}u$ flavour structures and have contributions from penguin operators as well as tree-level transitions; with respect to the group theory discussion, all irreps of the Hamiltonian can contribute, and specifically the penguin contributions are associated with the $\overline{\mathbf{3}}$ irrep.
In contrast, the CF $c\to s\bar{d}u$ and DCS $c\to d\bar{s}u$ transitions concern operators with four distinct flavours and live exclusively in the $\mathbf{6}\oplus\overline{\mathbf{15}}$ tree-level Hamiltonian.
While the group theory decomposition above highlights which \SUF-reduced matrix elements are most accessible, it is of interest to connect these to physical decay processes in terms of definite flavour states of specific CKM sectors.
Focusing on the $\overline{\mathbf{15}}\to\mathbf{27}$ transition, there are three physical decay amplitudes that are exclusively described by this in the \SUF\ limit:
\begin{align}
    \textbf{(CF)} ~~:~~ {\cal A}(D^+ \to \bar{K}^0\pi^+) &= \frac{V_{cs}^*V_{ud}}{\sqrt{2}}\Amp \,, \label{eq:DtoKpiCF} \\
    \textbf{(DCS)} ~~:~~ {\cal A}(D_s^+ \to K^0 K^+) &= \frac{V_{cd}^*V_{us}}{\sqrt{2}}\Amp \,,  \\
    \textbf{(SCS)} ~~:~~\, {\cal A}(D^+ \to \pi^0\pi^+) &= \frac{V_{cd}^*V_{ud}}{2}\Amp\,.
\end{align}
Similar discussions, including \SUF-breaking effects, can be found in e.g.~refs.~\cite{Hiller:2012xm,Bhattacharya:2021ndt}.
Note that while the CF and DCS components of the $\overline{\mathbf{15}}$ Hamiltonian are directly represented by operators with four distinct flavours, the SCS decay $D^+\to\pi^0\pi^+$ naturally repeats light flavours in the operator when considering definite flavour representations, however in the exact \SUF\ limit its relation to $\Amp$ must be understood as the symmetric, traceless $\overline{\mathbf{15}}$ projection.

\section{$(K\pi)_{\mathbf{27}}$ scattering amplitude}
\label{sec:kpi_scattering}
\begin{table}[th]
    \centering
    \begin{tabular}{c|cccccccc}
        label & $a~[\text{fm}]$ & $t_0/a^2$ & $\kappa_{ud}=\kappa_s$ & $(L/a)^3 \times T/a$ & $M_\pi L$ & $N_{\text{cfg}}$ & $N_{\rm vec}$ \\
        \hline
        {\tt a12m412} & 0.12  & 1.4868(04) & 0.1394305 & $24^3\times96$ & 6.01 & 77 & 60 \\
        {\tt a094m412} & 0.094 & 2.4400(01) & 0.138963  & $32^3\times96$ & 6.21 & 111 & 60 \\
        {\tt a064m412} & 0.064 & 5.2471(26)& 0.138272  & $48^3\times96$ & 6.40 & 74 & 60 \\
    \end{tabular}
    \caption{The OpenLat Initiative's \SUF-symmetric gauge-field ensembles with exponentiated Wilson-Clover fermions and the L\"uscher-Weisz gauge action used in this work~\cite{Luscher:2011bx,Francis:2019muy,Cuteri:2022erk,Cuteri:2022oms,Francis:2022hyr,Francis:2023gcm,Francis:2025pgf,Francis:2025rya}.
    Ensemble parameters are listed as determined by the OpenLat Initiative: lattice spacing $a$, Wilson-flow scale $t_0/a^2$, hopping parameter $\kappa_{ud}=\kappa_s$, lattice volume $(L/a)^3\times T/a$, and physical volume $M_\pi L$.
    The hopping parameters have been tuned to replicate~\cref{eq:SU3pion} using the CLS determination of the physical $t_0$~\cite{Bruno:2016plf}.
    The simulations performed here use $N_{\text{cfg}}$ independent configurations and $N_{\rm vec}$ Laplacian eigenvectors for distillation, as optimised in~\cite{Joswig:2022ctr}.}
    \label{tab:ensembles}
\end{table}

In the following, we discuss our computation of the $K\pi$ scattering amplitude in the \SUF\ 27-plet.
Our calculation is performed on the OpenLat Initiative's \SUF-symmetric ensembles listed in~\cref{tab:ensembles}.
The mass of the three degenerate quarks is tuned such that the trace of the mass matrix is equivalent to that of the physical world, i.e.
\begin{equation}
    m_u^{\rm phys} + m_d^{\rm phys} + m_s^{\rm phys} = 3m^{\rm latt}_\ell,
\end{equation}
or, equivalently in terms of hadron masses,
\begin{equation} \label{eq:SU3pion}
     M_K^2 = \frac13\left(2(M_K^{\rm phys})^2 + (M_\pi^{\rm phys})^2\right) \simeq (410.9\,{\rm MeV})^2\,,
\end{equation}
where $M_K$ refers, throughout, to the unphysical kaon (equal to the unphysical pion) mass of these ensembles.
We focus on the 27-plet, whose maximal-isospin $K\pi$ channel provides a particularly clean setting to determine the elastic scattering amplitude. 
The remainder of this section describes the construction and analysis of the correlation matrices used to obtain the finite-volume spectrum: \cref{sec:operators} introduces the distillation framework and interpolating operators; \cref{sec:two_point} gives the resulting correlation functions; \cref{sec:kpi_spectrum} details the GEVP analysis; and \cref{sec:kpi_phase} determines the scattering phase information from the finite-volume spectrum both at finite lattice spacing and in the continuum.

\subsection{Distillation framework and operator construction}
\label{sec:operators}
Determining correlation functions for processes involving multi-hadron states comes at a large computational cost since, to capture the dynamics of the internal quarks within these scattering hadrons, `all-to-all' propagators are typically required.
A well-established method in hadronic spectroscopy to reduce the cost associated with propagator creation is distillation~\cite{HadSpec:2009krc,Morningstar:2011ka}.
Distillation is a technique to project propagators into a small subspace of their huge size and subsequently reduce the cost of computation.
Using stout-smeared~\cite{Morningstar:2003gk} gauge fields $U_\mu(\vx,t)$, one considers from the lattice Laplacian,
\begin{equation}
    -\nabla^2_{nm}(\vx,\vy;t) = 6\delta_{\vx,\vy}\delta_{nm} - \sum_{j=1}^3\left(U_j^{nm}(\vx,t)\delta_{\vx+\hat{j},\vy} + U_j^{mn}(\vx-\hat{j},t)^*\delta_{\vx-\hat{j},\vy}\right).
\end{equation}
We are interested in the lowest-energy states of hadrons constructed from these quark fields, and expect that the dominant contributions for these will come from the lowest-lying modes of the Laplacian's spectrum.
Obtaining these modes requires solving the eigenvalue equation 
\begin{equation}
    -\sum_\vy \nabla^2(\vx,\vy;t) v_k(\vy,t) = \lambda_k(t)v_k(\vx,t),
\end{equation}
where a cut-off parameter must be chosen to make this affordable. 
This cut-off is the number of eigenvectors used, $N_{\rm vec}$ and $v_k(\vx,t)$ are then the lowest $N_{\rm vec}$ eigenvectors of the Laplacian subspace.
The {\it distillation operator} is a form of smearing kernel and is defined 
\begin{equation}
    {\cal S}(\vx,\vy;t) = \sum_{k=1}^{N_{\rm vec}} v_k(\vx,t)v_k(\vy,t)^\dagger \equiv V(\vx,t)V^\dagger(\vy,t).
\end{equation}
The dimension of $V(\vx,t)$ is $N_c\times N_T\times N_L^3\times N_{\rm vec}$ for $N_c$ colours and $N_L$ lattice sites in each spatial direction.
This is much smaller than the full unsmeared space ($(N_c\times N_T\times N_L^3)^2$) which would amount to millions of Laplacian eigenvalues. 

We can also apply dilution projectors
\begin{equation}
    P^{[d]} = \delta_{tt'}\,\delta_{\alpha\beta}\,\delta_{kl}
\end{equation}
to the distillation setup which projects time slices $t'$, spin indices $\beta$, and Laplacian eigenvectors $l$ into a compound dilution index $[d] = [t',\beta,l]$.
While more general dilution projectors can be used to define a stochastic distillation~\cite{Morningstar:2003gk}, here we use {\it exact distillation} through Kronecker deltas to project one index triplet $(t,\alpha,k)$ for each compound index $[d]$.

Applying the distillation operator to both quark fields in a propagator, one finds
\begin{equation} \label{eq:DistilProp}
    \begin{aligned}
        {\cal S}\,D^{-1}\,{\cal S} &= V\,V^\dagger\,D^{-1}\,V\,V^\dagger \\
                                   &= \sum_{[d]} V\,V^\dagger D^{-1}\,V\,P^{[d]}\,P^{[d]\dagger}\,V^\dagger \\
                                   &= \sum_{[d]} V\,\tau^{[d]}\,\left[P^{[d]\dagger}\,V^\dagger\right]
    \end{aligned}
\end{equation}
where we may identify the {\it perambulator} 
\begin{equation} \label{eq:DistilPeramb}
    \tau^{(f)[d]}(t;t') = \sum_{\vx,\vy} V(\vx,t)^\dagger\,D^{-1}_{f}(\vx,t;\vy,t')\,\left[V(\vy,t')\,P^{[d]}\right],
\end{equation}
with $f$ denoting the quark flavour.\footnote{Note that throughout this section, we set $a=1$ and thus do not explicitly show factors of $a$.}
Furthermore, it is convenient to factorise the propagator into a {\it source vector},
\begin{equation} \label{eq:DistilRho}
    \varrho^{[d]}(\vx,t) = V(\vx,t)\,P^{[d]},
\end{equation}
and a {\it sink vector} 
\begin{equation} \label{eq:DistilPhi}
    \varphi^{(f)[d]}(\vx,t;t') = V(\vx,t)\,\tau^{(f)[d]}(t;t').
\end{equation}
It is convenient to assemble these into $(4N_TN_{\rm vec})^2$ matrices known as {\it meson fields},
\begin{equation} \label{eq:MesonField}
    M_{\Gamma,\nu\omega}^{[d_1,d_2]}(\vp,t) = \sum_\vx e^{-i\vp\cdot\vx}\,{\rm Tr}\left[\nu^{[d_1]}(\vx,t)^\dagger\,\Gamma\,\omega^{[d_2]}(\vx,t)\right],
\end{equation}
where $\nu,\omega\in\{\varrho,\varphi\}$, $\Gamma$ is some Dirac structure, and the trace is performed over spin and colour.
The construction so far is useful when studying spectral quantities, however the nature of distillation as a smearing operation means that in this form it cannot be used to form local operator insertions for e.g.\ weak decays.
While~\cref{eq:DistilRho} is no different in principle to other smeared-source techniques, the challenge is in the sink vector and thus the perambulator, which is projected into LapH space at both ends of its quark line.
To ensure the quark fields entering a local operator insertion are truly local, we can first invert the Dirac matrix before projecting at the sink as is done in~\cref{eq:DistilPeramb}, and write the original smeared perambulator as
\begin{equation}
    \tau^{(f)[d]}(t;t') = \sum_{\vx} V(\vx,t)^\dagger \phi^{(f)[d]}(\vx,t;t'),
\end{equation}
where $\phi$ is the {\it unsmeared} or {\it generalised} perambulator and still contains the fully local information of the quark line at position $t$ while only position $t'$ has been projected to LapH space.\footnote{Techniques such as {\it blending} may allow for the reduction of cost associated with such objects~\cite{Hu:2025vhd}.}
A generic local bilinear operator with Dirac structure $\Gamma$ can then be constructed as
\begin{equation} \label{eq:localBil}
    B_{\Gamma,ab}^{(f_1,f_2)[d_1,d_2]}(\vx,t;t_1,t_2) = \sum_{\alpha\beta} \phi_{a\alpha}^{(f_1)[d_1]}(\vx,t;t_1)^\dagger\,\Gamma_{\alpha\beta}\,\phi_{b\beta}^{(f_2)[d_2]}(\vx,t;t_2),
\end{equation}
where $a,\,b$ are colour indices.
We will apply this to the construction of the weak decay in~\cref{sec:dkpi_decay}.

\subsection{Two-point correlation functions}
\label{sec:two_point}
Single-hadron interpolating operators for the \SUF\ pion can be constructed with definite momentum,
\begin{equation} \label{eq:pi_interp}
    O_\pi(\vp;t) = \sum_{\vx} e^{-i\vp\cdot\vx} \,\bar{\ell}(\vx,t)\,\gamma_5\, \ell(\vx,t),
\end{equation}
where $\ell$ is the \SUF\ light quark and the momenta to be considered are
\begin{equation} \label{eq:mom_lim}
    \vp = \frac{2\pi}{L}\vd_\pi, \quad \vd_\pi \in\mathbb{Z}^3, \quad 0\leq\vd_\pi^2\leq4.
\end{equation}
Using the distillation framework, these pion interpolators are realised via meson fields $M_{\gamma_5,\varrho\varphi}^{[d_1,d_2]}(\vp,t)$ in~\cref{eq:MesonField} from which pion correlation functions can be constructed as
\begin{equation}
    C_\pi(\vp_\pi;t,t') = -\sum_{d_1,d_2} M_{\gamma_5,\varrho\varphi}^{[d_1,d_2]}(\vp_\pi; t)\,M_{\gamma_5,\varrho\varphi}^{[d_2,d_1]}(-\vp_\pi; t').
\end{equation}
Two-particle interpolating operators are then formed from combinations of~\cref{eq:pi_interp} satisfying a fixed total momentum,
\begin{equation}
    \vP = \vp_K + \vp_\pi,
\end{equation}
according to
\begin{equation}
    O_{K\pi}(\vp_K,\vp_\pi;t) = O_K(\vp_K;t)\,O_\pi(\vp_\pi;t).
\end{equation}
Note that we use the labels $K$ and $\pi$ to refer equivalently to the interpolating operator of~\cref{eq:pi_interp}; as discussed in~\cref{sec:group_theory}, the maximal-isospin channel of interest here projects to $K\pi$-like states from the \SUF\ pseudoscalar octet.

In infinite volume, multi-particle states are classified according to the irreducible representations (irreps) of the rotation group SO(3), labelled by angular momentum $(l,m)$.
On a finite cubic lattice, however, rotational symmetry is reduced to the cubic symmetry group.
Consequently, the operator basis is projected onto irreps of the appropriate lattice symmetry group.
For zero total momentum the relevant symmetry group is the octahedral group $O_h$, while for non-zero total momentum the little group that leaves $\vP$ invariant,
\begin{equation}
    R\vP = \vP
\end{equation}
is used.
The projection onto an irrep $\Lambda$ is performed using the standard group-theoretical projection operator,
\begin{equation}
    O^{\Lambda}= \frac{d_{\Lambda}}{|G|} \sum_{R\in G}\Gamma^{(\Lambda)}(R)^{*}\,U(R)\,O,
\end{equation}
where $G$ denotes the relevant symmetry group, $d_{\Lambda}$ is the dimension of the irrep, $\Gamma^{(\Lambda)}(R)$ is the representation matrix (or character for one-dimensional irreps), and $U(R)$ applies the symmetry operation $R$ to the operator basis by rotating the momentum components.
Since only one-dimensional irreducible representations are considered in this work, namely $A_1^+$ in the rest frame and $A_1$ in moving frames, the projection reduces to a sum over the characters of the corresponding irrep,
\begin{equation}
    O^{\Lambda} = \frac{1}{|G|} \sum_{R\in G} \chi^{(\Lambda)}(R)^*\,U(R)\,O.
\end{equation}
The projected operators transform according to a definite lattice irrep and lead to correlation matrices that are block diagonal in $\Lambda$.
These projected correlation matrices will subsequently be used in a variational analysis to determine the finite-volume energy spectrum.

For the present study, the $s$-wave $K\pi$ scattering channel is accessed through the $A_1^+$ irrep of the octahedral group $O_h$ in the rest frame and the $A_1$ irrep of the relevant little groups in moving frames.
The $A_1^+$ irrep contains continuum partial waves with $l=0,4,6,\ldots$, while the moving-frame $A_1$ irreps contain contributions from multiple partial waves due to the reduced lattice symmetry, however the spectrum is dominated by the $l=0$ contributions.

\begin{figure}[t]
    \centering
    \includegraphics[width=\textwidth]{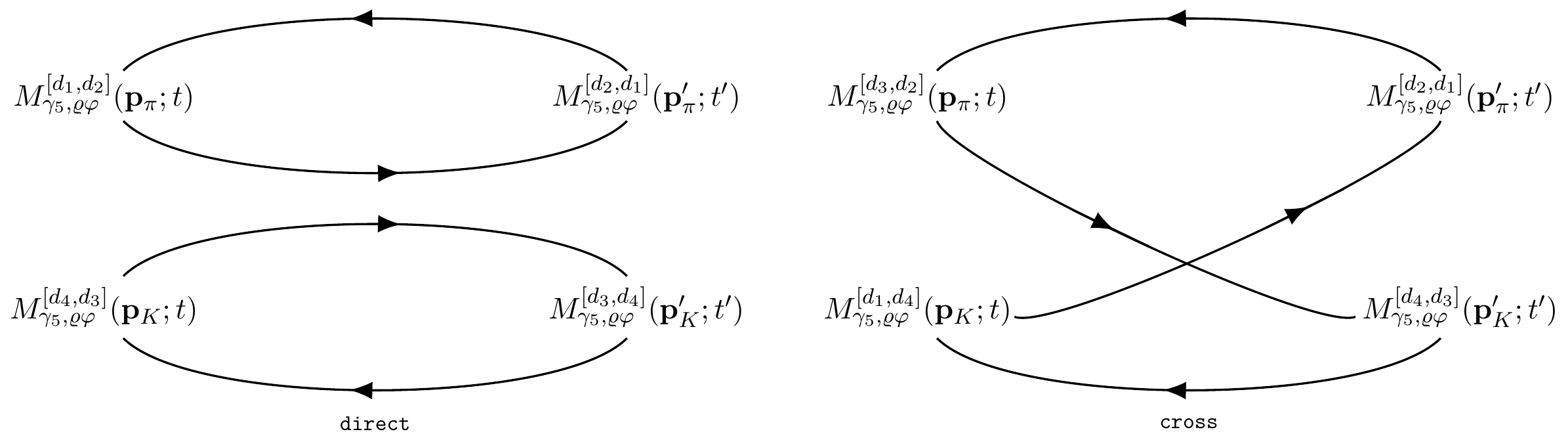}
    \caption{Wick contractions contributing to maximal-isospin $K\pi$ scattering.}
    \label{fig:Wick2pt}
\end{figure}

With all this in mind, we again use meson fields to efficiently construct the $2\to2$ $K\pi$ scattering correlation functions
\begin{equation}
    C_{K\pi}(\vp_K,\vp_\pi,\vp_K',\vp_\pi';t,t') = \langle O_{K\pi}(\vp_K,\vp_\pi;t)\,O_{K\pi}(\vp_K',\vp_\pi';t')^\dagger\rangle.
\end{equation}
The scattering correlation function has multiple Wick contractions contributing, namely the `{\tt direct}' and `{\tt cross}' diagrams which are shown in~\cref{fig:Wick2pt}.
Up to global normalisation, the correlator of the maximal-isospin channel is expressed as
\begin{equation} \label{eq:2ptCorr}
    C_{K\pi}(\vp_K,\vp_\pi,\vp_K',\vp_\pi';t,t') = C_{K\pi}^{\tt direct} - C_{K\pi}^{\tt cross}.
\end{equation}
We compute~\cref{eq:2ptCorr} for all momenta in~\cref{eq:mom_lim} and all $N_T$ time source positions, subsequently translating and averaging across these source positions; without loss of generality, from here on we will refer to the source position $t'=0$.
In addition, we identify the correlators projected to equivalent total momenta from the little group symmetries and average these as well.

\subsection{Finite-volume spectrum}
\label{sec:kpi_spectrum}
Using the correlators constructed above, we can form a $n_{\rm op}\times n_{\rm op}$ {\it correlator matrix}
\begin{equation}
    C_{IJ}^{\vP}(t,t_0) \equiv \langle O_I^{\vP}(t)\,O_J^{\vP}(0)^\dagger\rangle
\end{equation}
in each total momentum frame $\vP$ where $I,\,J$ span the $n_{\rm op}$ interpolators projecting to this irrep and frame; in this study, since e.g.\ there are no resonances present, this amounts simply to different momentum choices distributed between the two particles in the $K\pi$ state.
Solving the {\it generalised eigenvalue problem} (GEVP)~\cite{Luscher:1990ck,Blossier:2009kd,Fischer:2020bgv},
\begin{equation}
    C^{\vP}(t) u_n^{\vP}(t,t_0) = \lambda_n^{\vP}(t,t_0)C^{\vP}(t_0)u_n^{\vP}(t,t_0)
\end{equation}
to yield eigenvalues $\lambda_n^{\vP}(t,t_0)$ and their associated eigenvectors $u^\vP_n(t,t_0)$ with a reference time $t_0$ providing variationally optimised linear combinations of the interpolating operator basis that have maximal overlap with individual finite-volume spectral energies.
From here on, we will omit the dependence on $t_0$ in the eigenvalues and eigenvectors, though they are still assumed.

After solving the GEVP on our correlator matrices, we extract the finite-volume energies by fitting to the asymptotic form of the eigenvalues,
\begin{equation} \label{eq:EVcorr}
    \lambda_n^\vP(t) = Z_n^\vP\,e^{-E_n^\vP\,t}\,\left[1+{\cal O}\left(e^{-\Delta_n^\vP\,t}\right)\right],
\end{equation}
where $E_n^\vP$ is the $n^{\rm th}$ finite-volume energy level of the momentum frame $\vP$ and $\Delta_n^\vP$ describes residual contamination from excited states.
We perform two independent analyses to extract the finite-volume energy levels from the eigenvalue correlators:
\begin{itemize}[topsep=1pt,itemsep=1pt]
    \item {\bf Analysis 1} performs combined correlated fits between all energy levels in a given momentum frame using the model-averaging procedure described in ref.~\cite{Jay:2020jkz}.
    Using $t_0\in\{2,3\}$, each energy level is fitted to the leading term of~\cref{eq:EVcorr} for regions in Euclidean time $[t_{n,{\rm min}}^\vP,t_{n,{\rm max}}^\vP]$ and we consider all possible sequential series in these windows with 2 or more degrees of freedom per energy level. 
    \Cref{tab:gevp-energies-a12,tab:gevp-energies-a094,tab:gevp-energies-a064} in appendix~\ref{app:GEVPtables} show the time windows and fit results for each finite-volume energy level.
    \item In {\bf Analysis 2}, we perform individual correlated one- and two-exponential fits for all allowed $t_{n,\rm{min}}^{\vP}$ values at a given $t_{n,{\rm max}}^{\vP}$ and a model average is performed as described in ref.~\cite{Jay:2020jkz}. 
    We assess the stability of the spectrum by varying $t_0$ from $2$ to $12$ and include multiple $t_0$ values for each eigenvalue which keep the fitted energy stable.
    The chosen $t_{n,\mathrm{max}}^{\vP}\leq 2t_0$ ensures that the GEVP is evaluated in the regime $t_0\geq t/2$, for which the excited state contamination is asymptotically suppressed by the energy gap to the first state omitted from the variational basis, $\mathcal O\left(e^{-\Delta_n^{\vP}t}\right)$~\cite{Blossier:2009kd}.
\end{itemize}
The resulting energy levels of the two analyses are found to be in excellent agreement with one another. The fitted energy levels normalised to the non-interacting two-particle levels are shown in~\cref{fig:GEVP_nonintRAT_P0} for the rest frame; similar plots for the moving frames are shown in appendix~\ref{app:GEVPtables}. 

The finite temporal extent induces thermal wrap-around contributions to the $K\pi$ correlators. 
The leading thermal effects are proportional to $e^{-E_\pi(\boldsymbol q) T}e^{-(E_K(\boldsymbol p)-E_\pi(\boldsymbol q))t}$ (analogously with $K\leftrightarrow\pi$). Since $M_K = M_\pi$ at the \SUF-symmetric point, whenever $\boldsymbol p = \boldsymbol q$, these terms are $t$-independent and thus act as a constant shift on the correlation function: $C(t) = A e^0 + B e^{- E_0 t} + \cdots$ where $E_0$ is the lowest lying state in the $T \to \infty$ correlator and $A$ and $B$ are overlaps, with $B$ suppressed by $e^{- E_\pi(\boldsymbol q)T}$. The wrap-around effect therefore manifests as an effective vacuum contamination. For intermediate $t$, the term is numerically negligible due to the suppressed overlap, so the correlator decays with the energy of interest, $E_0$, but for sufficiently large $t$ the correlator reveals the lower-lying state. In an effective mass plot, this manifests as an intermediate plateau at $E_0$ that then gives way to the true zero-energy plateau; see~\cref{fig:GEVP_nonintRAT_P0}.  
We verify that this does not affect the extracted spectrum by repeating the analysis using the finite-difference correlator, $ C(t)-C(t+1)$, to remove the leading constant contribution~\cite{Dudek:2012gj}. The resulting energy levels are consistent with those from the original analysis.

\begin{figure}[th]
    \centering
    \includegraphics[width=0.32\textwidth]{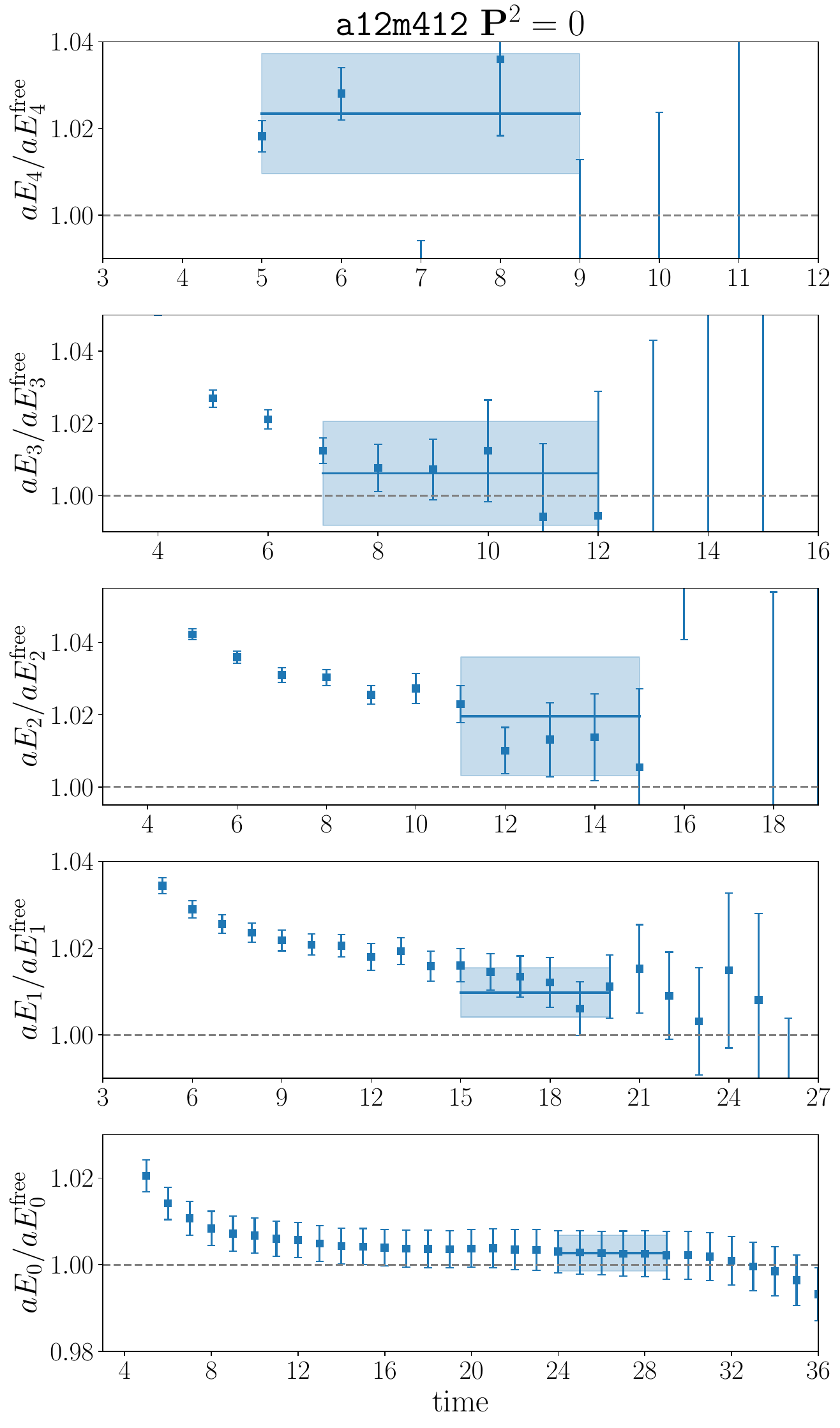}
    \includegraphics[width=0.32\textwidth]{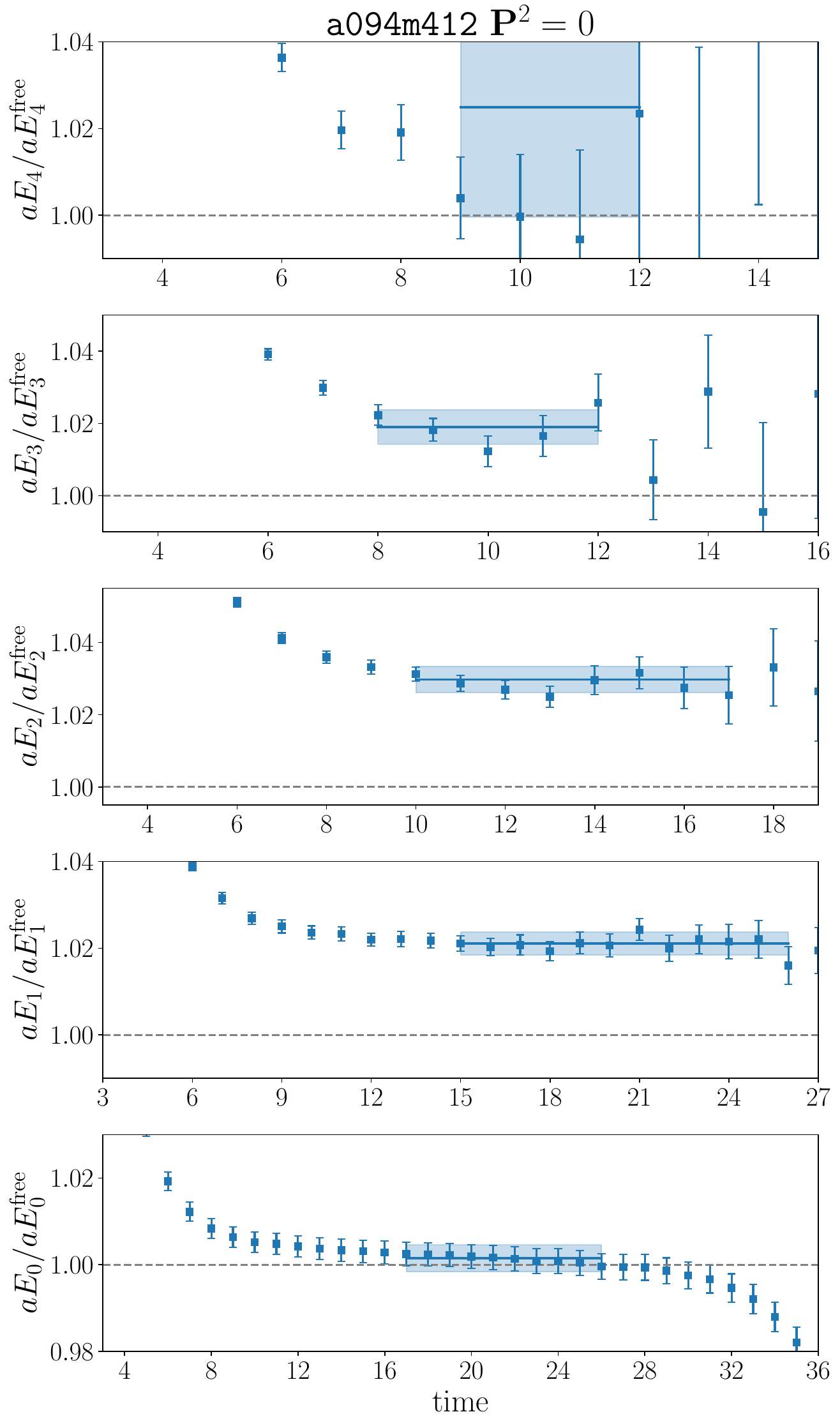}
    \includegraphics[width=0.32\textwidth]{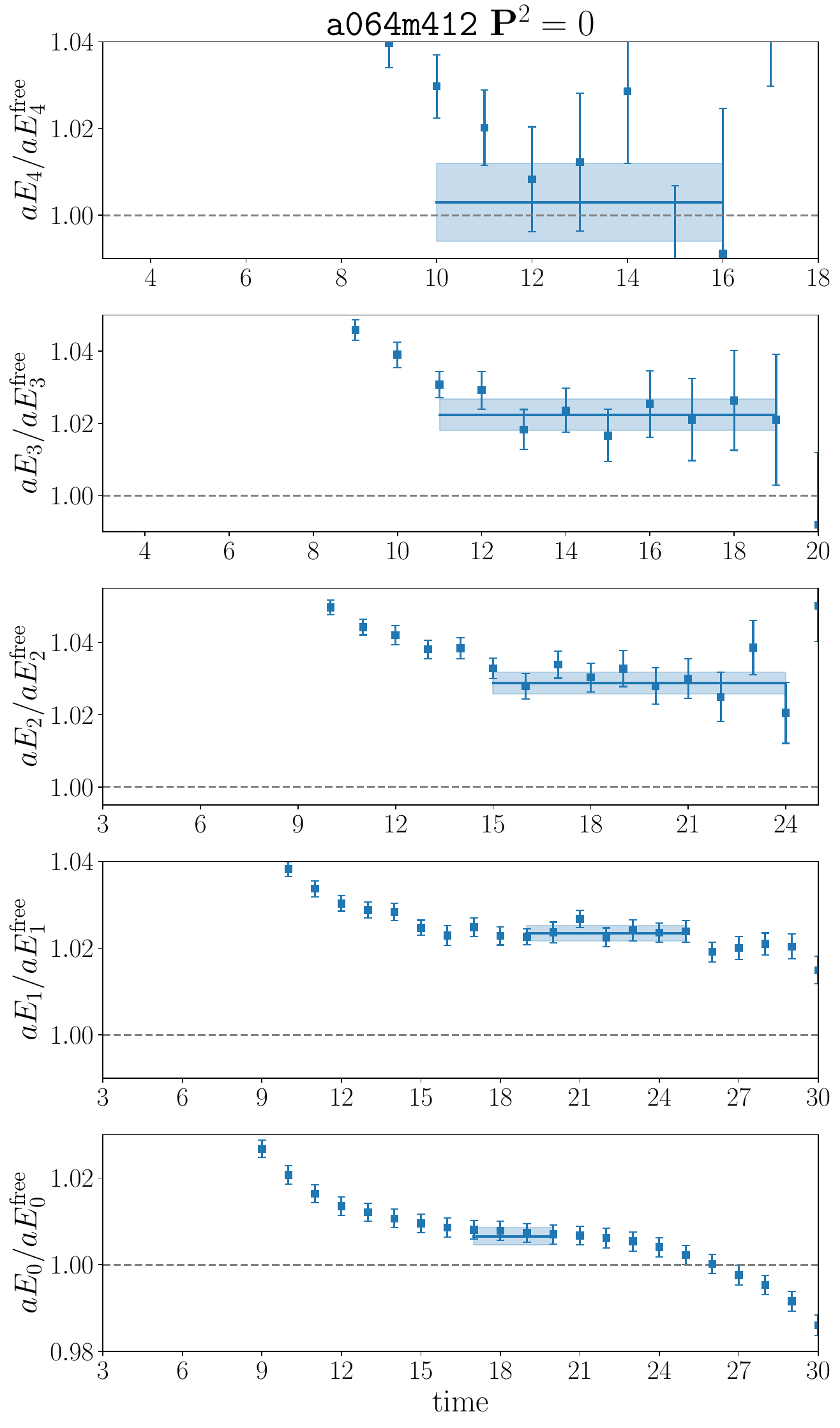}
    \caption{Effective energies of the GEVP eigenvalues $\lambda_n^\vP(t)$ and their fitted finite-volume energy levels normalised to the non-interacting energy levels indicated by the dashed lines for the $\vP^2=0$ frame. The {\tt a12m412} ensemble is shown in the left column, the {\tt a094m412} in the middle, and {\tt a064m412} on the right.}
    \label{fig:GEVP_nonintRAT_P0}
\end{figure}

\subsection{Phase shift extraction}
\label{sec:kpi_phase}
The general paradigm relevant to this work is that finite-volume data, accessible in a numerical lattice QCD calculation, can be related to physical, infinite-volume scattering and decay amplitudes.
Before focusing on the specific simplified framework used in this calculation, we recall the most general approach.
For any system with $N$ open flavour channels at a centre-of-mass-frame energy $E_{\sf cm}$, one can define a K matrix, $\mathcal K$, that encodes the scattering dynamics at that energy.
The finite-volume spectrum is then obtained from a quantisation condition of the schematic form
\begin{equation}
    \det\left[ \mathcal K(E_{\sf cm})^{-1} + F(E,\vP,L;\mathcal K^{\rm sub})\right] = 0 \,,
    \label{eq:generic_QC}
\end{equation}
where $\mathcal K$ is a matrix acting on the combined flavour-channel space and on the index space describing angular momentum and, where relevant, other kinematic labels of the open channels.
The finite-volume matrix $F$ encodes the box size $L$, the total momentum $\vP$ in the finite-volume frame, and the scattering of subprocesses indicated by the $\mathcal K^{\rm sub}$ argument.
This quantisation condition is satisfied for each finite-volume energy, and lattice-determined energies can therefore be used to constrain $\mathcal K$.
The latter is then used in on-shell integral equations to determine the physical scattering amplitudes, schematically
\begin{equation}
    \mathcal M = \mathcal I[\mathcal K,\mathcal K^{\rm sub}] \,.
    \label{eq:generic_integral_equation}
\end{equation}
This approach is well established for two-particle channels, beginning with L\"uscher's original work and its extensions to moving frames, coupled channels and arbitrary spin~\cite{Luscher:1986pf,Luscher:1990ux,Rummukainen:1995vs,Hansen:2012tf,Briceno:2012yi,Briceno:2014oea,Wilson:2015dqa}.
Analogous formalisms have also been developed for three-particle systems and for coupled two- and three-particle channels~\cite{Hansen:2014eka,Briceno:2017tce}, with perturbative and threshold-expansion results available for more general $N$-particle systems~\cite{Romero-Lopez:2020rdq,Mukherjee:2026fourpion}.

\begin{figure}[th]
    \centering
    \includegraphics[width=0.322\textwidth]{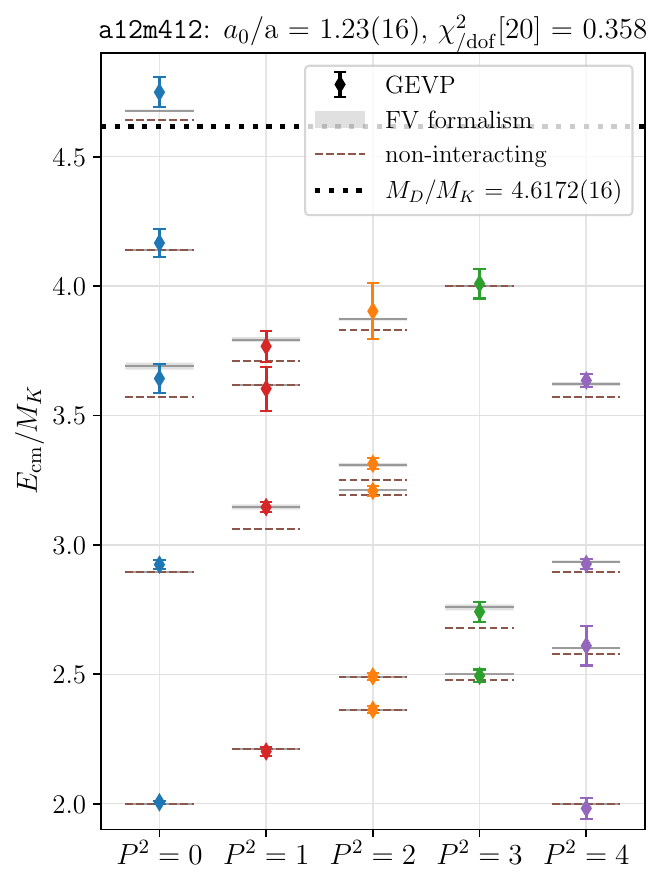}
    \includegraphics[width=0.33\textwidth]{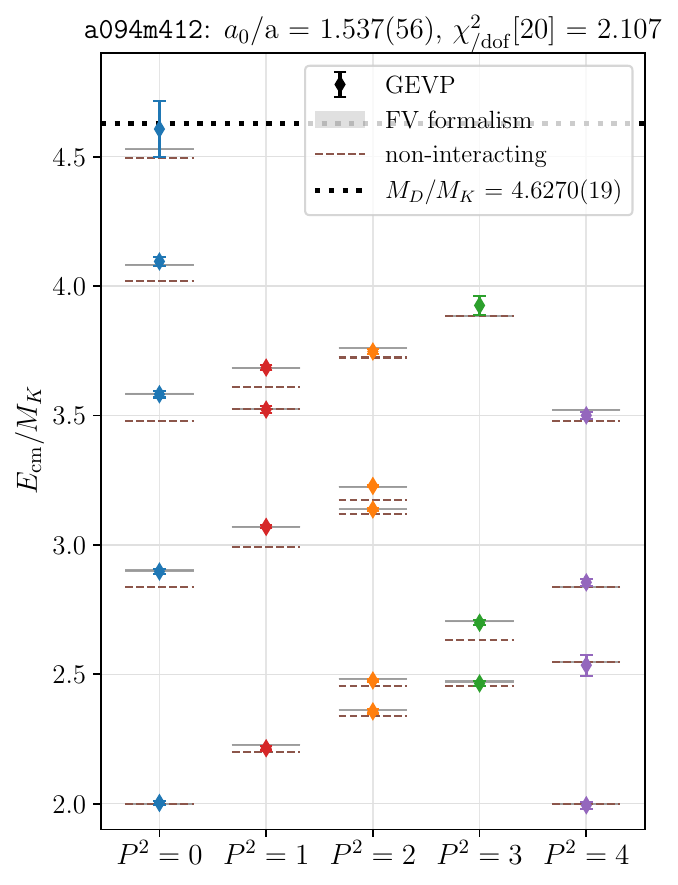}
    \includegraphics[width=0.33\textwidth]{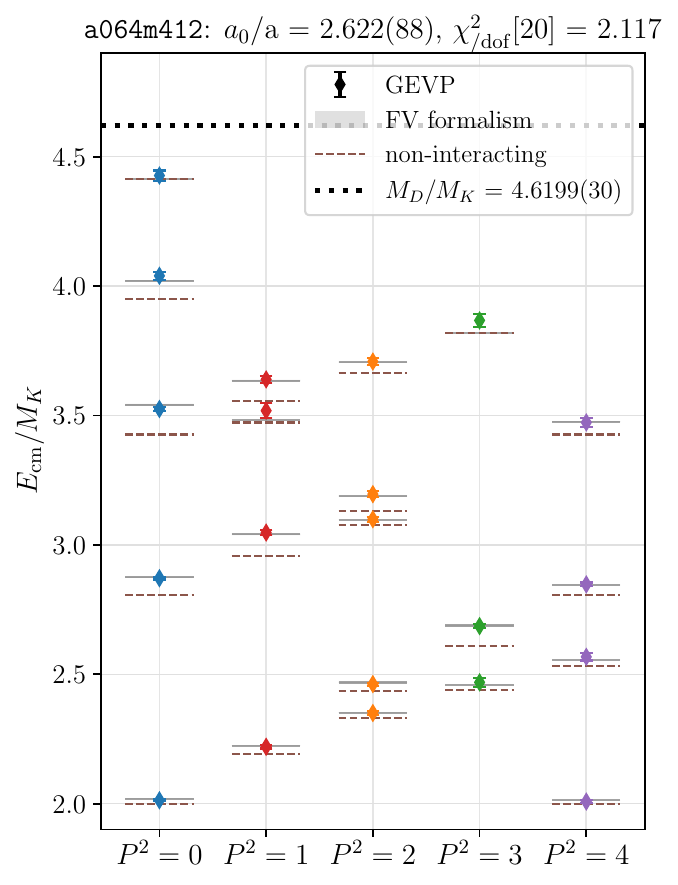}
    \caption{Finite-volume energy levels on the {\tt a12m412} (left), {\tt a094m412} (centre), {\tt a064m412} (right) ensembles.
        The coloured data show the values extracted from the GEVP while the gray bands are derived from the fit of the SL phase shift parametrisation (\cref{eq:pcot_sL}) propagated through the finite-volume quantisation condition. The brown dashed lines indicate the non-interacting two-particle energy levels and the black dotted line shows the mass of the \SUF-averaged $D$ meson.}
    \label{fig:scatteringlengthColumns}
\end{figure}

In the present case we have a single 27-plet two-particle channel with $J^P=0^+$, so a three-pseudoscalar state cannot couple in the $s$-wave since it has negative parity. 
The next coupled channel therefore arises only for $E_{\sf cm} > 4M_\pi$.
We neglect the effect of this four-particle threshold as it is only relevant for a few of the highest-lying finite-volume energies that we extract (see, for example,~\cref{fig:scatteringlengthColumns}) and the associated systematic uncertainty is expected to be small.

Thus the quantisation condition reduces to that of a single two-particle channel first worked out by L\"uscher in refs.~\cite{Luscher:1986pf,Luscher:1990ux}.
We additionally restrict attention to the trivial irreducible representation of either the octahedral group for $\vP=0$, or of the associated little group for $\vP\neq0$.
This means that the matrices entering the quantisation condition must be projected to that irrep.
Even after this projection, it is still true formally that an infinite tower of angular-momentum channels contributes to each finite-volume energy.

The lowest-lying contribution is the $s$ wave, with vanishing angular momentum $l=0$, and the next contribution after this is the $g$ wave with $l=4$.
The latter is expected to be highly suppressed both by the weak interactions of the \SUF\ hadrons in the 27-plet and by the angular-momentum barrier, which leads to a $p^{2l}$ suppression for sufficiently small scattering momentum $p$.
We therefore additionally take the approximation of truncating the K matrix to only its $s$-wave component.
In this approximation only a single component of $\mathcal K$ contributes and the determinant reduces to the simple algebraic result
\begin{equation}
    p\cot\delta_0(p) =
    \frac{2}{\gamma L\sqrt{\pi}}\,
    \mathcal Z_{00}^{\vd}\left(\left(\frac{pL}{2\pi}\right)^2\right) \,,
    \label{eq:single_channel_luscher}
\end{equation}
where $p$ is the scattering momentum in the centre-of-mass frame, $E_{\sf cm} = 2 \sqrt{p^2 + M_\pi^2}$, $\delta_0$ is the $s$-wave phase shift, $\gamma=E/E_{\sf cm}$ is the boost factor, and $\vd=\vP L/(2\pi)$.
Here we have used the relation between $F$ and the L\"uscher zeta function, which for identical scattering particles is defined as $ \mathcal Z_{00}^{\vd}(x^2) =  \mathcal Z_{00}^{\vd}(1;x^2)$, with
\begin{equation}
    \mathcal Z_{00}^{\vd}(s;x^2)
    =
    \frac{1}{\sqrt{4\pi}}
    \sum_{\mathbf r\in P_{\vd}}
    \frac{1}{(\mathbf r^2-x^2)^s} \,,
    \qquad
    P_{\vd}
    =
    \left\{
    \mathbf r=\hat{\gamma}^{-1}\left(\mathbf n-\frac{1}{2}\vd\right)
    \, \middle| \,
    \mathbf n\in\mathbb Z^3
    \right\},
    \label{eq:zeta_definition}
\end{equation}
where $\hat{\gamma}^{-1}$ acts on a vector by multiplying the component parallel to $\vP$ by $\gamma^{-1}$ while leaving the perpendicular components unchanged.

In practice, we consider various parametrisations of the scattering phase shift,
\begin{equation}
    \delta_0(p) = \delta_0^{\sf par}(p,\boldsymbol{\alpha}) \,,
\end{equation}
where ${\sf par}$ labels a particular parametrisation and $\boldsymbol{\alpha}$ is a vector of parameters.
The weakly repulsive 27-plet system is well described by simple polynomials in $p^2$, and in particular we consider the following two possibilities:
\begin{align}
    p\cot\delta_0^{\sf SL}(p)
    &=
    -\frac{1}{a_0} \,,
    \label{eq:pcot_sL}
    \\
    p\cot\delta_0^{\sf ERE}(p)
    &=
    -\frac{1}{a_0}
    + \frac{1}{2} r_0 p^2 \,,
    \label{eq:pcot_ERE}
\end{align}
where $a_0$ is the scattering length and $r_0$ is the effective range.
Each parametrisation leads to a prediction for the trivial-irrep energies in a given moving frame as a function of the input parameters,
\begin{equation}
    E_{L,n,\vd}^{\sf par}(\boldsymbol{\alpha}) \,,
\end{equation}
where $L$ labels the spatial extent of the finite volume, $n$ labels the finite-volume level, and $\vd$ the spatial momentum in the finite-volume frame.
Combining these predictions with the lattice data, one can then form a chi-squared,
\begin{equation}
    \chi^2_{\sf par}(\boldsymbol{\alpha})
    =
    \sum_{i,j}
    \left[
    E_i^{\rm latt}
    -
    E_i^{\sf par}(\boldsymbol{\alpha})
    \right]
    \left[C^{-1}\right]_{ij}
    \left[
    E_j^{\rm latt}
    -
    E_j^{\sf par}(\boldsymbol{\alpha})
    \right] \,,
    \label{eq:qfit_chisq}
\end{equation}
where the compound index $i=(L,n,\vd)$ runs over all included finite-volume energies and $C$ is the corresponding covariance matrix.
The best-fit parameters are then determined by minimising~\cref{eq:qfit_chisq} for each parametrisation separately.
We find that $r_0$ remains consistent with zero while the resulting $a_0$ across both ans\"atze are very consistent for all ensembles. 
We also consider ans\"atze including an Adler zero term which is also consistent with zero while $a_0$ remains the same within uncertainties.
As a further cross-check that the results are not overly impacted by the less reliable highest energy states from the GEVP (cf.\ the upper rows of~\cref{fig:GEVP_nonintRAT_P0}) whose values could be infected by e.g.\ 4-particle contamination, we repeat each of the phase shift analyses with a cut on the energy at $E_{\rm cm}=3.7M_K$; once again we do not find any significant deviations in these results.
Therefore we take the SL phase-shift determination using~\cref{eq:pcot_sL} as our central values.
The finite-volume energy spectrum for each ensemble using the determined SL phase shifts are compared to the levels resulting from the GEVP in~\cref{fig:scatteringlengthColumns}. 

Absent from this discussion so far is the role of discretisation effects, i.e.\ lattice-spacing artefacts, on the finite-volume energy spectrum.
Strictly, the formalism described above is only applicable to continuum finite-volume energies.
Therefore any analysis performed on a given ensemble implicitly assumes that discretisation effects are either subdominant to statistical uncertainties or else can be absorbed into the K-matrix parameters.
The effects of lattice artefacts on the quantisation condition were investigated in ref.~\cite{Hansen:2024discretization}.
There, a somewhat involved proposal is given for including the discretisation effects on the single-hadron dispersion relation into a modified form of the L\"uscher zeta function.
On the dataset used in this work, however, we do not resolve statistically significant lattice artefacts on the boosted single-pion energies.
Thus, if these effects are statistically negligible, it is justified to incorporate lattice artefacts in the parameters $\boldsymbol{\alpha}$.
We consider this approach below.

\subsubsection{Scattering length continuum limit}
\label{sec:contlim}

\begin{figure}[th]
    \centering
    \includegraphics[width=0.48\textwidth]{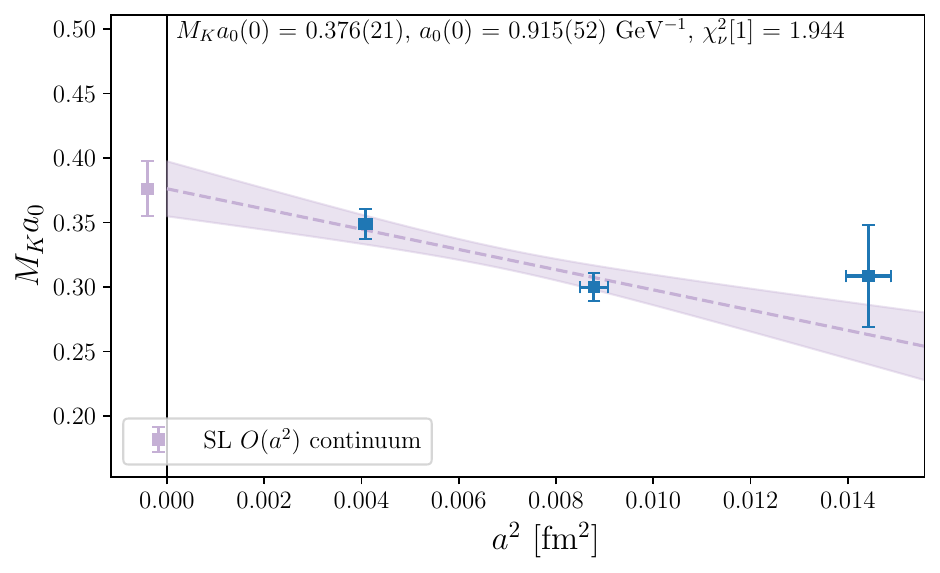}
    \includegraphics[width=0.48\textwidth]{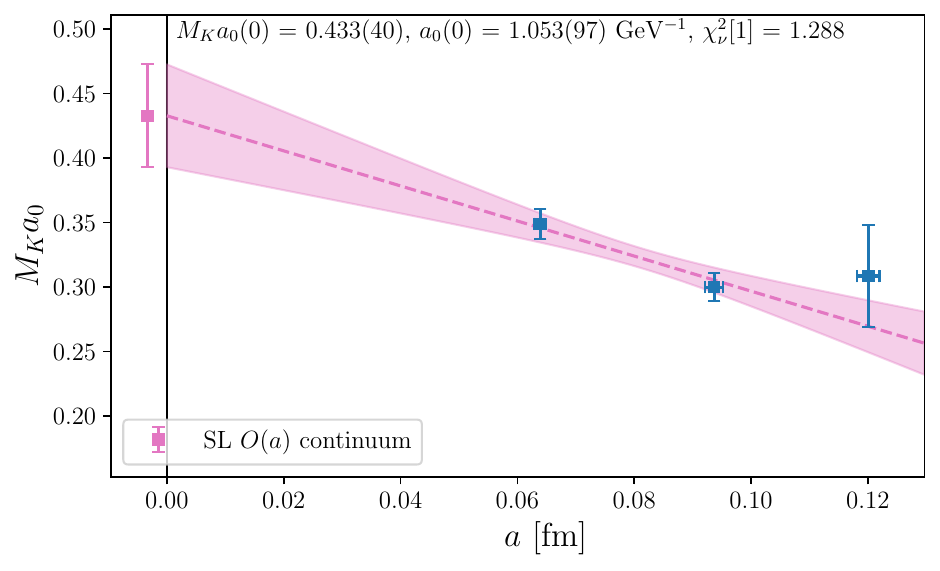}
    \caption{Continuum extrapolations of the $I=3/2$ \SUF\ $K\pi$ scattering length using both $O(a^2)$ (left) and $O(a)$ (right) ans\"atze.}
    \label{fig:continuum_ERE}
\end{figure}

For each lattice spacing, the scattering length $a_0/a$ is obtained as described above.
To compare the results, we can form the dimensionless quantity $M_K a_0$; since the three ensembles in~\cref{tab:ensembles} are tuned to the same \SUF\ mass, this quantity is then identical up to residual discretisation effects.
Spectral quantities are $O(a)$ improved for the exponentiated Wilson-clover action and thus our central ansatz for a continuum extrapolation is linear in $a^2$:
\begin{equation}
    M_Ka_0(a) = M_Ka_0(a=0) + C\,a^2.
\end{equation}
As a conservative check of the assumed scaling, we also repeat the extrapolation for a linear-in-$a$ ansatz.
The resulting fits are shown in~\cref{fig:continuum_ERE}.
While there is a slight trend downwards in the $O(a^2)$ extrapolation compared to $O(a)$, the two results are still compatible within uncertainties.
We maintain our choice of the $O(a^2)$ extrapolation as our central result and include the comparison with the $O(a)$ result in our measure of the systematic uncertainty below.

\subsubsection{Finite-volume energies in the continuum}
\label{sec:continuumFVEs}

\begin{figure}[th]
    \centering
    \includegraphics[width=\textwidth]{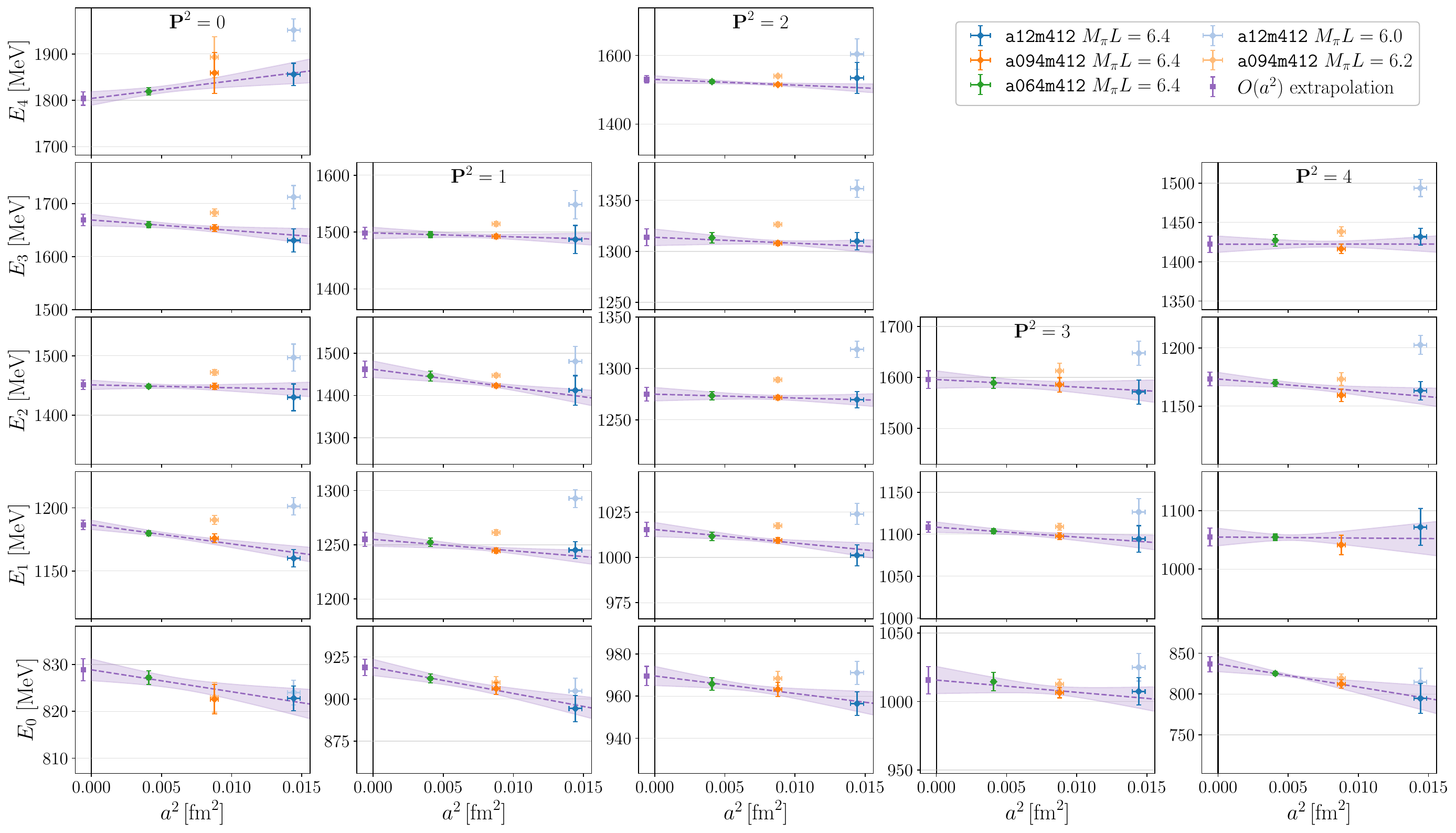}
    \caption{Continuum results of the finite-volume energy spectra. Each momentum frame is shown in its own column from the rest frame $\vP^2=0$ on the left to $\vP^2=4$ on the right. 
    The original data of the two smaller-volume ensembles are shown in faint colours, while the values corrected to $M_\pi L=6.4$ are shown in solid colours. 
    The continuum extrapolations are given in purple by the dashed line and faint uncertainty band. The final continuum result is indicated in the left of each panel.}
    \label{fig:continuum_FVEs}
\end{figure}

The discrete energy spectrum itself is a product of the finite volume, and can be considered in the continuum as well as at finite lattice spacing.
This is appealing because the energies are the primary lattice observables, while infinite-volume scattering parameters are derived nonlinearly through the finite-volume formalism.
Therefore as an alternative approach, we take the continuum limit of each finite-volume energy at fixed physical volume --- a method that is perhaps more rigorous and also gives higher confidence on the overall consistency of our dataset.
To do so, we need to match to a common physical volume.
The three ensembles are already reasonably well tuned to this aim, with a maximum variation of $M_\pi L$ of approximately $6\%$; see~\cref{tab:ensembles}.
To correct this small mistuning, we take our dedicated L\"uscher analysis on each ensemble and use the resulting scattering parameters to determine the expected shift induced by the mistuning of the volume.
Defining $L_{\rm target}$ as the physical volume of the finest ensemble, {\tt a064m412}, we can evaluate the shift
\begin{equation}
    E_{L_{\rm target},n,\vd}^{\sf par}(\boldsymbol{\alpha})
    -
    E_{L,n,\vd}^{\sf par}(\boldsymbol{\alpha}) \,,
    \label{eq:volume_shift}
\end{equation}
and add this to the lattice-determined energy.
For the weakly-interacting system we are considering, the finite-volume expansion of the energies about their non-interacting levels has been derived in ref.~\cite{Grabowska:2021xkp} in terms of the scattering length, and therefore the required volume correction in~\cref{eq:volume_shift} can be computed using this definition.
This then gives a three-lattice-spacing trajectory for each energy at fixed physical volume, to which one can apply an $a\to0$ extrapolation.
As discussed above in~\cref{sec:contlim}, these quantities are $O(a)$ improved and therefore we perform the continuum extrapolation using a linear ansatz in $a^2$.
For a consistency check, we also perform the extrapolation using a linear ansatz in $a$ and find good agreement.
The fits are performed using dimensionless energies in terms of the fixed \SUF\ pion mass, i.e.\ $E_n/M_K$.
The continuum extrapolations of the finite-volume energies at $M_\pi\,L\simeq6.4$ are shown in~\cref{fig:continuum_FVEs}, where the fainter colours indicate the original finite-volume energies on the {\tt a12m412} and {\tt a094m412} before the correction of ref.~\cite{Grabowska:2021xkp}.
The left panels show the continuum results of both ans\"atze.
The results and associated $\chi^2$ and $p$-values are listed in~\cref{tab:cl-energies-comparison} in appendix~\ref{app:CLtables}.

\begin{figure}[th]
    \centering
    \includegraphics[width=0.55\textwidth]{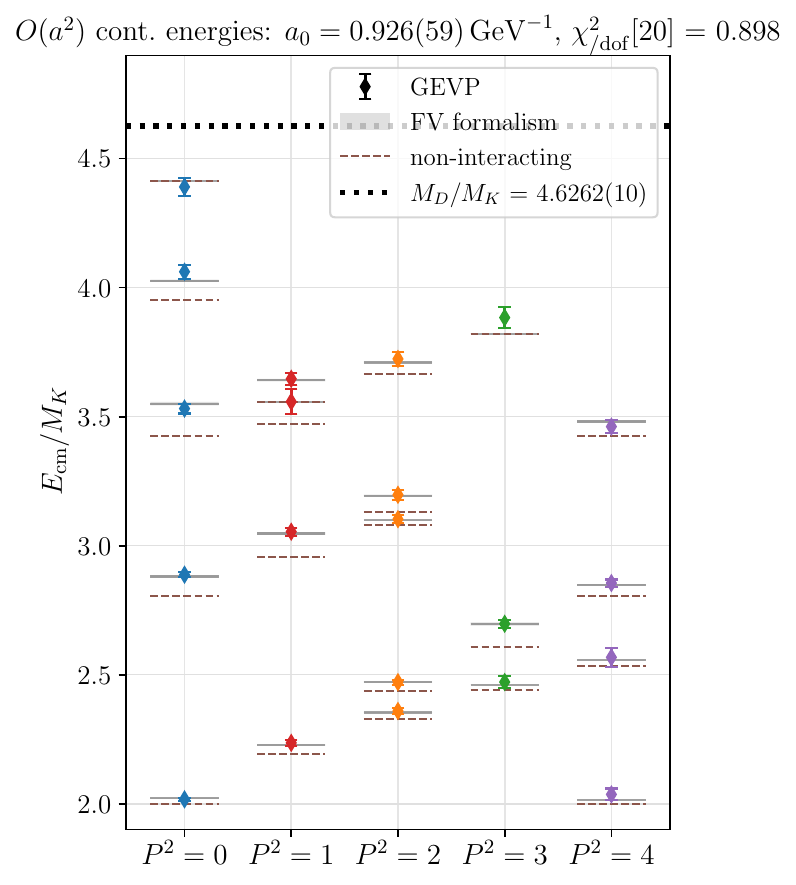}
    \caption{Finite-volume energy levels in the continuum from an $O(a^2)$ extrapolation.
        The coloured data show the values extracted from the GEVP and continuum fits while the gray bands are derived from the fit of the SL phase shift parametrisation (\cref{eq:pcot_sL}) propagated through the finite-volume quantisation condition. The brown dashed lines indicate the non-interacting two-particle energy levels and the black dotted line shows the mass of the \SUF-averaged $D$ meson.}
    \label{fig:continuum_a0}
\end{figure}

\begin{figure}[th]
    \centering
    \includegraphics[width=0.8\textwidth]{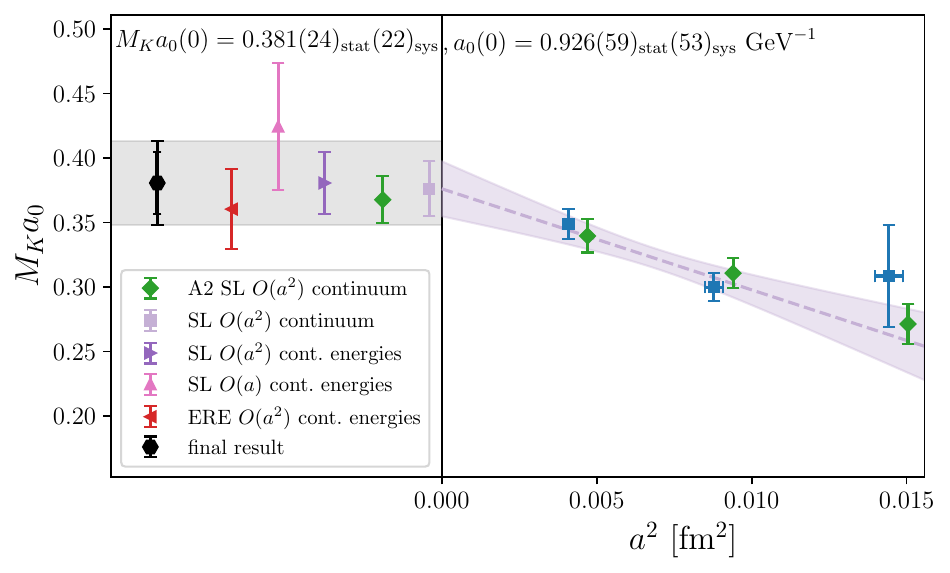}
    \caption{Final result for the $I=3/2$ \SUF\ $K\pi$ scattering length given by the black data point, given by the SL result using $O(a^2)$-extrapolated finite-volume energies (purple) with the systematic uncertainty included. 
        Alternative results using continuum finite-volume energies are shown by the pink and red coloured points respectively using the SL ansatz with $O(a)$-extrapolated energies and the ERE ansatz with $O(a^2)$-extrapolated energies.
        The faint point in the left panel shows the $O(a^2)$ continuum extrapolation of the scattering length itself (using the SL phase-shift determination), and the green data point shows the result of Analysis 2 for the same choice. 
        The right panel shows the $O(a^2)$ continuum extrapolation where the individual ensemble results are shown in blue for Analysis 1 (cf.~\cref{fig:continuum_ERE}) and in green for Analysis 2 with a small horizontal offset.}
    \label{fig:continuum_ERE_compare}
\end{figure}

Using the continuum finite-volume energy levels, the phase shift analysis described above can be repeated to arrive directly at the continuum result for the scattering length.
We consider both ans\"atze of~\cref{eq:pcot_sL,eq:pcot_ERE} and once again find consistency in the resulting values of $a_0$ while further parameters are compatible with zero.
The continuum finite-volume energy spectrum using the determined SL phase shift is compared to the levels resulting from the $O(a^2)$ extrapolation in~\cref{fig:continuum_a0}. 
Finally, these results are compared with those obtained by extrapolating the scattering length itself in~\cref{fig:continuum_ERE_compare}. 
The two procedures agree perfectly well with one another.
We take the result in~\cref{fig:continuum_a0} obtained from the $O(a^2)$ extrapolation of the finite-volume energies using the SL phase shift as our central value. 
To estimate the systematic uncertainty in our result, we compare half-differences of our central value and the central values of: a) the $O(a^2)$-extrapolated SL phase shift from the individual ensemble results, b) the $O(a^2)$-extrapolated SL phase shift of Analysis 2 (A2) c) the SL phase shift with $O(a)$-extrapolated continuum energies, and d) the ERE phase shift with $O(a^2)$-extrapolated continuum energies.
The final systematic uncertainty is taken as the maximum of these four alternative results.
As one can see in the figure, all these variations are well compatible with our central choice and do not significantly increase the total uncertainty. 
Our final result for the scattering length of the $I=3/2$ \SUF\ $K\pi$ system is
\begin{equation}
    a_0 = 0.926(59)_{\rm stat}(53)_{\rm sys}\,{\rm GeV}^{-1} = 0.926(79)\,{\rm GeV}^{-1}.
\end{equation}

In the elastic, single-channel approximation, unitarity implies that, after factoring out the CKM weak phase, the strong phase of the $\Amp$ decay amplitude is given by the $({K\pi})_{\mathbf{27}}$ scattering phase shift, $\delta_{\mathbf{27}}(E)$, modulo $\pi$. 
To illustrate the size of this phase at the \SUF-averaged $D$-meson mass, we evaluate the central SL parametrisation at $E=M_D$:
\begin{equation} \label{eq:delta27}
    \delta_{\mathbf{27}}(E=M_D) = -38.4(2.4)^\circ.
\end{equation}
This phase shift fixes the strong phase associated with the $\Amp$ decay amplitude.
Although an overall phase of a decay amplitude is convention-dependent, relative strong phases between contributions to a physical decay amplitude are observable. 
The determination of $\delta_{\mathbf{27}}$ therefore provides a necessary first ingredient for predicting such relative phases once the corresponding amplitudes and final-state interactions in the remaining \SUF\ channels are determined.

\section{$D \to (K\pi)_{\mathbf{27}}$ decay}
\label{sec:dkpi_decay}

\subsection{Weak Hamiltonian and operator renormalisation}
\label{sec:weak_operators}
We start with the effective weak Hamiltonian
\begin{equation}
    {\cal H}_W = \frac{G_F}{\sqrt{2}}\left\{\lambda_d\left[C_1(\mu)Q_1^d + C_2(\mu)Q_2^d\right] + \lambda_s\left[C_1(\mu)Q_1^s + C_2(\mu)Q_2^s\right] + \lambda_b\sum_{i\geq 3}C_i(\mu)Q_i\right\},
\end{equation}
with CKM coefficients $\lambda_x = V_{cx}V_{ux}^*$ and the tree-level operators
\begin{align}
    Q_1^q &= \bar{q}^\alpha\gamma_\mu(1-\gamma_5)c^\beta \otimes \bar{u}^\beta\gamma^\mu(1-\gamma_5)q^\alpha, \label{eq:effH_Q1} \\
    Q_2^q &= \bar{q}^\alpha\gamma_\mu(1-\gamma_5)c^\alpha \otimes \bar{u}^\beta\gamma^\mu(1-\gamma_5)q^\beta, \label{eq:effH_Q2}
\end{align}
where $\alpha,\beta$ are the colour indices.
$Q_{i\geq3}$ are the penguin operators with the heavy $b$ quark and $W$ boson integrated out; we will not consider these any further here.

The $\Amp$ amplitude of interest in this study is a tree-level decay and thus only the two operators $Q_1$ and $Q_2$ contribute.
The Dirac structures of the two operators are identical, but $Q_1$ is a colour-rearranged operator and $Q_2$ a colour singlet.
While the operators have $(\mathrm{V}-\mathrm{A}) \otimes (\mathrm{V}-\mathrm{A})$ structure, parity limits which components are non-zero.
Since the $D$ meson has $P=-1$ and the $s$-wave $K\pi$ state has $P=+1$, then the non-zero components of the four-quark operators must have $P=-1$, i.e. we can focus on the ${\rm V}\otimes{\rm A}+{\rm A}\otimes{\rm V}$ sub-structures of $Q_{1,2}$.
From here on when referring to $Q_{1,2}$, we will assume the negative-parity components.

\subsection{Three-point correlation function setup}
\label{sec:three_point}

Following the group theory discussion of~\cref{sec:group_theory}, we have identified the $\Amp$ amplitude as a convenient first target.
This corresponds to e.g.\ the Cabibbo-favoured decay $D^+\to\bar{K}^0\pi^+$ in~\cref{eq:DtoKpiCF} which proceeds via two ``flavour-flow'' topologies -- the `colour-favoured tree' (T) and `colour-suppressed tree' (C) diagrams.
The additional `weak annihilation' (A), `weak exchange' (E), and `singlet weak exchange' (SE) diagrams enter in the classification of all amplitudes, but do not contribute to $\Amp$; see e.g.~\cite{Ryd:2009uf}.
From the different diagrams and operators, we can identify four total contributions to $\Amp$: $T^{Q_1},\,T^{Q_2},\,C^{Q_1},\,C^{Q_2}$.
The amplitude of a specific operator $\Amp^{Q_i}$ is constructed from the individual diagrams following
\begin{equation}
    \frac{1}{\sqrt{2}}\Amp^{Q_i} = T^{Q_i} + C^{Q_i}.
\end{equation}

We will compute the T and C diagrams in our distillation framework where source vectors and smeared sink vectors have already been assembled into meson fields for the scattering analysis in~\cref{sec:kpi_scattering}, however there is the additional requirement to calculate unsmeared perambulators $\phi$ to form the local four-quark operator.

\begin{figure}[th]
    \centering
    \includegraphics[width=0.6\textwidth]{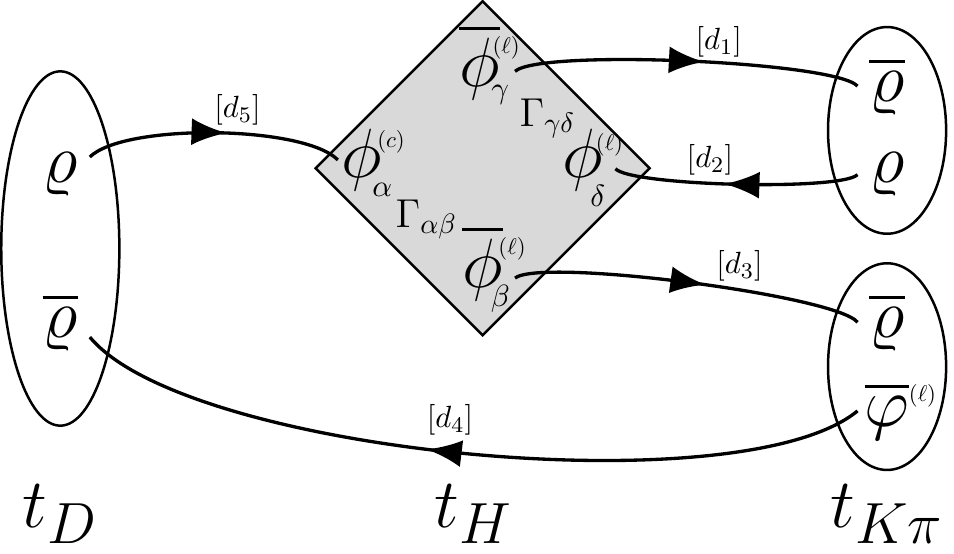}
    \caption{T diagram for the $D\to K\pi$ decay in terms of the distillation source and sink vectors. Meson fields representing the $D$ meson and $K\pi$ final state are placed at times $t_D$ and $t_{K\pi}$ respectively and the four-quark operator of the effective Hamiltonian is inserted at time $t_H$ using unsmeared perambulators. Distillation indices $[d_i]$ shared across quark lines, quark flavours $(q)$, and Dirac indices $\alpha,\beta,\gamma,\delta$ are shown explicitly.}
    \label{fig:DtoKpiTDistil}
\end{figure}

\begin{figure}[th]
    \centering
    \includegraphics[width=0.6\textwidth]{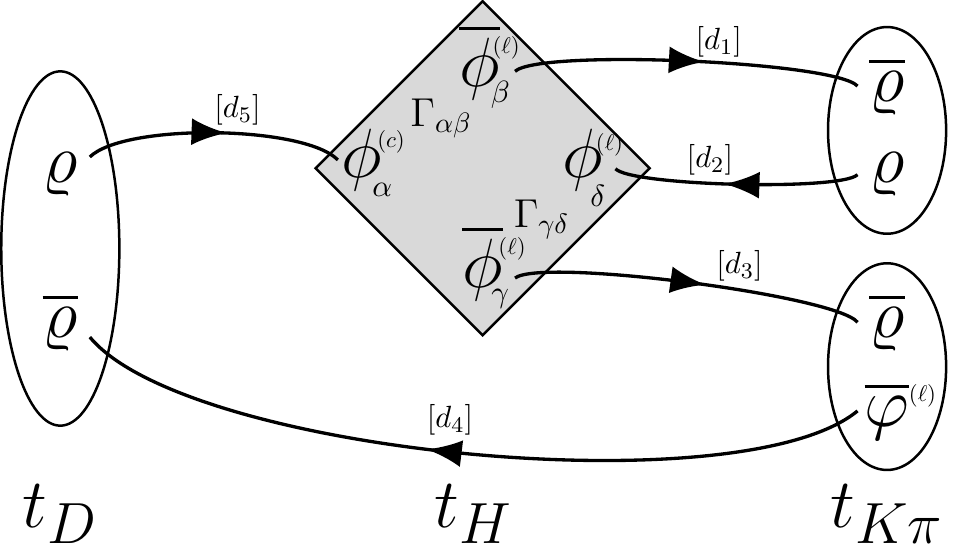}
    \caption{C diagram for the $D\to K\pi$ decay in terms of the distillation source and sink vectors. Meson fields representing the $D$ meson and $K\pi$ final state are placed at times $t_D$ and $t_{K\pi}$ respectively and the four-quark operator of the effective Hamiltonian is inserted at time $t_H$ using unsmeared perambulators. Distillation indices $[d_i]$ shared across quark lines, quark flavours $(q)$, and Dirac indices $\alpha,\beta,\gamma,\delta$ are shown explicitly.}
    \label{fig:DtoKpiCDistil}
\end{figure}

In~\cref{fig:DtoKpiTDistil,fig:DtoKpiCDistil}, we respectively show the T and C diagrams with explicit distillation objects and indices shown.
These diagrams contain an insertion of a four-quark current.
Since the weak Hamiltonian is a local operator, the corresponding four-quark current must also be local.
As described above, this can be accommodated within the distillation framework by modifying the usual construction such that quark lines are smeared at only one end, rather than at both.
This requires the computation of generalised perambulators. In \cref{eq:localBil} we showed how these can be used to construct a local meson field.
The same idea extends naturally to four-quark operators.
There are two main ingredients.
First, in analogy to the usual meson field, we can define a four-quark field by combining four distillation vectors and projecting the resulting object onto a definite momentum $\vp$.
Second, by constructing this object from generalised perambulators, the four-quark operator can be made fully local at the position of the weak Hamiltonian.

This construction is possible provided that no propagator both starts and ends at the four-quark operator.
Such a propagator would have to be local at both ends and therefore cannot be constructed using the one-sided generalised perambulators employed here.
It would instead require an additional estimation technique.
These self-contractions are absent for our computation in the 27-plet, so the four-quark operator can be treated entirely within distillation.

The principal difficulty is instead the size of the four-quark field itself.
To illustrate this, consider first the storage required for an ordinary meson field in our setup.
Its size scales as $(T/a)^2 \cdot (N_s \cdot N_v)^2 \cdot N_{\vp} \cdot N_\Gamma \cdot 8\, \mathrm{bytes} = 96^2 \cdot 240^2 \cdot 33 \cdot 2 \cdot 8\, \mathrm{bytes} = 260\, \mathrm{GB}$ per configuration.
Here we use $N_{\vp}=33$ momentum rotations, corresponding to all momentum rotations up to $\vd^2=4$, and the two (pseudo)scalar Dirac structures $\Gamma \in {\mathds{1},\gamma_5}$.
For the four-quark fields, a larger set of Dirac structures is required.
In particular, we require eight combinations corresponding to $VA$ and $AV$, with $V\in{\gamma_x,\gamma_y,\gamma_z,\gamma_t}, A=V\gamma_5$.
On the other hand, only a single momentum is required, since the weak Hamiltonian $H_W$ is always projected to zero momentum.
The requirement that $H_W$ remain local means that the final spatial sum must be performed only after constructing the operator at each local insertion point.
We can, however, reduce the number of source times used in constructing the meson fields and the subsequent three-point functions.
Rather than using all 96 time slices, we use 24 source times.
Even with this reduction, a naive construction of the four-quark field would require $96 \cdot 24 \cdot 240^4 \cdot 1 \cdot 8 \cdot 8\, \mathrm{bytes}= 495\, \mathrm{TB}$ per configuration.
Storing and manipulating such an object is impractical for our calculation.
We therefore avoid constructing the complete four-quark field explicitly and instead directly assemble the three-point function from smaller intermediate objects.
In practice, we construct these intermediates from the meson fields associated with the $D$ meson and the two mesons forming the $K\pi$ final state, together with the relevant generalised perambulators.
This allows the necessary contractions to be performed while summing over subsets of the distillation indices during contraction, thereby avoiding the prohibitively large storage requirement of the full four-quark field.

The T diagram in~\cref{fig:DtoKpiTDistil} effectively factorises into the product of two Dirac-distillation traces.
We can therefore compute the two intermediate objects for this diagram:
\begin{align}
    X_{\Gamma,ab}(\vp_{K\pi};\vx_H,t_H,t_{K\pi}) &= \sum_{d_1,d_2} B_{\Gamma,ab}^{(\ell,\ell)[d_2,d_1]}(\vx_H,t_H)\,M^{[d_1,d_2]}_{\gamma_5,\varrho\overline{\varrho}}(\vp_{K\pi};t_{K\pi}) \label{eq:Xab} \\
    Y_{\Gamma,cd}(\vp_D,\vq_{K\pi};\vx_H,t_D,t_H,t_{K\pi}) &= \sum_{d_3,d_4,d_5} \bigg[M^{[d_4,d_3]}_{\gamma_5,\overline{\varphi}\,\overline\varrho}(\vq_{K\pi};t_{K\pi}) \, B_{\Gamma,cd}^{(\ell,c)[d_3,d_5]}(\vx_H,t_H) \notag \\
    &\qquad\qquad\qquad\qquad\qquad\qquad\quad\times M^{[d_5,d_4]}_{\gamma_5,\varrho\varphi}(\vp_{D};t_{D})\bigg], \label{eq:Ycd}
\end{align}
where $\Gamma_1,\,\Gamma_2$ are the Dirac structures of the two bilinears composing the four-quark operator and $a,\,b,\,c,\,d$ are colour indices.
Here, $B$ is the local bilinear meson field defined in \cref{eq:localBil}. Crucially, this object has not yet been projected onto a definite momentum and is therefore itself large.
By carrying out the contractions over the available dilution indices at this intermediate stage, however, the overall computation remains tractable.
Note that we have suppressed the source-time indices on $B$ as they are implicit through the distillation index trace.

These intermediate objects then allow for efficient computation of the T diagram for both $Q_1$ and $Q_2$ via
\begin{align}
    T^{Q_1}(\vp_D,\vp_{K\pi},\vq_{K\pi};t_D,t_H,t_{K\pi}) &= \sum_{\vx_H} {\rm Tr}\left[X_{\Gamma_1,ab}(\vx_H)\, Y_{\Gamma_2,ba}(\vx_H)\right] \\
    T^{Q_2}(\vp_D,\vp_{K\pi},\vq_{K\pi};t_D,t_H,t_{K\pi}) &= \sum_{\vx_H} {\rm Tr}\left[X_{\Gamma_1,ab}(\vx_H)\right]\times {\rm Tr}\left[Y_{\Gamma_2,cd}(\vx_H)\right],
\end{align}
where the traces are over colour indices and we have abbreviated the dependence of $X_{\Gamma,ab}$ and $Y_{\Gamma,cd}$ on momenta and time positions shown in~\cref{eq:Xab,eq:Ycd}.
This is computationally favourable as the factorised sums of $d_1,d_2$ and $d_3,d_4,d_5$ allow for the reuse of indices.

While~\cref{fig:DtoKpiCDistil} can also be computed directly for the C diagram, it is naively more expensive than the T diagram since it has a single non-factorisable sum over all distillation indices.
However, using Fierz relations it is possible to express the $Q_1$ and $Q_2$ C contributions in terms of T diagrams with different linear combinations of Dirac matrices and opposite colour structure.
Specifically,
\begin{equation}
    C^{Q_1/Q_2} = T_{\rm AV-VA}^{\rm s/r},
\end{equation}
where s and r indicate colour-singlet and rearranged contractions respectively.
Therefore by computing the T diagram for all variations of $VA$ and $AV$ four-quark Dirac structures for both colour contractions, we can form different linear combinations to reach all four of $T^{Q_1},\,T^{Q_2},\,C^{Q_1},\,C^{Q_2}$ which are needed for $\Amp$.

Finally, we remark that similar strategies are possible for other tree-level decay topologies (see e.g.~ref.~\cite{Bhattacharya:2021ndt}), and in fact, while they will be associated with further complexity in the group theory decomposition, ``exchange'' and ``annihilation'' topologies are computationally cheaper since no quark line goes directly from the initial to final state and thus the diagrams can be factorised across $t_H$.
These diagrams will be explored further in the future when moving beyond $\Amp$.

\subsection{Lellouch--L\"uscher formalism}
\label{sec:ll_formalism}

We now come to the task of relating the renormalised finite-volume matrix elements to the physical infinite-volume decay amplitudes.
This can be achieved using the formalism of Lellouch and L\"uscher~\cite{Lellouch:2000pv}, together with extensions to the case in which multiple two-particle channels are open~\cite{Kim:2005gf,Hansen:2012tf,Briceno:2014uqa}.
As with the quantisation condition above, we begin by describing the general framework before specialising to the simplified case of a single two-particle channel.
The general framework begins with the matrix entering the quantisation condition in~\cref{eq:generic_QC}.
For convenience we define
\begin{equation}
    \mathcal Q(E,\vP,L)
    =
    \mathcal K(E_{\sf cm})
    +
    F(E,\vP,L;\mathcal K^{\rm sub})^{-1} \,,
    \label{eq:ll_qc_matrix}
\end{equation}
so that the finite-volume energies satisfy $\det\mathcal Q(E,\vP,L)=0$.
Following ref.~\cite{Briceno:2014uqa}, the corresponding Lellouch--L\"uscher matrix is given by the residue of the inverse quantisation-condition matrix at a finite-volume pole,
\begin{equation}
    \mathcal R(E_n,\vP,L)
    =
    \lim_{E\to E_n}
    (E-E_n)\,
    \mathcal Q(E,\vP,L)^{-1} \,.
    \label{eq:ll_residue_matrix}
\end{equation}
This matrix acts on the same combined channel, angular-momentum and kinematic index space as $\mathcal K$ and $F$.
Given an infinite-volume transition-amplitude vector $\mathcal A$, with components labelled by the allowed on-shell two-particle channels and partial waves, the finite-volume matrix element extracted from a three-point function is related to $\mathcal A$ by
\begin{equation}
    \left|\langle E_n,\vP,L | \mathcal H_W(0) | D,\vP \rangle_L\right|^2
    =
    \frac{1}{2E_D}\,
    \mathcal A^\dagger
    \mathcal R(E_n,\vP,L)
    \mathcal A \,,
    \label{eq:ll_master_relation}
\end{equation}
up to convention-dependent normalisation factors and exponentially suppressed finite-volume effects.
The key point is that the same scattering information used to constrain the finite-volume spectrum also fixes the proportionality between finite-volume matrix elements and physical decay amplitudes.

For the present calculation, the same simplifications discussed in the scattering analysis reduce this matrix relation to a scalar: only a single elastic two-particle channel is retained and the amplitude is truncated to the $s$ wave.
The residue factor can then be expressed in terms of the scattering phase shift $\delta_0$ and the finite-volume pseudo-phase $\phi$.
With the normalisation conventions used here, the relation becomes
\begin{equation}
    |{\cal A}|^2 = \big|{\cal C}^{\rm LL}\big|^2 \left|\langle E_n,\vP,L | \mathcal H_W(0) | D,\vP \rangle_L\right|^2,
\end{equation}
where
\begin{equation} \label{eq:CLL}
    \big|{\cal C}^{\rm LL}\big|^2 = 8\pi\left[q\frac{\partial\phi}{\partial q}+p\frac{\partial\delta_0}{\partial p}\right]_{p=p_n}\frac{E^2_n\,M_D}{p^3_n}.
\end{equation}
Here $p_n$ is the centre-of-mass relative momentum corresponding to the finite-volume energy level $E_n$, $q=pL/2\pi$ is the dimensionless momentum, and $\phi(q)$ is the kinematic pseudo-phase appearing in L\"uscher's quantisation condition.
Thus, in addition to the finite-volume matrix element, the final finite-volume amplitude requires the energy dependence of the $s$-wave phase shift.
The derivative of $\delta_0$ is evaluated using the SL phase-shift parametrisation determined in~\cref{sec:kpi_scattering}.

\begin{figure}[th]
    \centering
    \includegraphics[width=\textwidth]{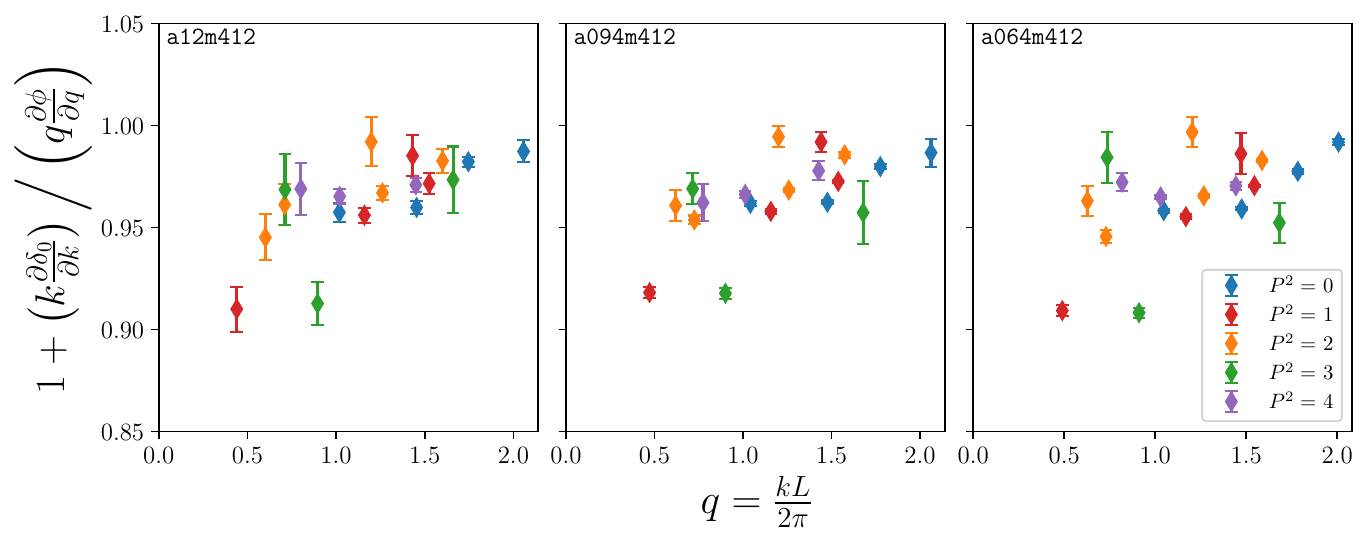}
    \caption{Ratios of the Lellouch--L\"uscher factor to its non-interacting value for the {\tt a12m412} (left), {\tt a094m412} (middle), and {\tt a064m412} (right) ensembles using the SL parameterisation of the phase shift.}
    \label{fig:LLratios}
\end{figure}

Using the phase-shift information determined in~\cref{sec:kpi_scattering}, we have determined the Lellouch--L\"uscher factors in~\cref{eq:CLL} for the SL phase-shift parametrisation. 
The determined values, and their associated relative momenta, are listed in~\cref{tab:LLfactors} for all finite-volume energy levels on each ensemble.
To illustrate the size of the interaction-dependent contribution, in~\cref{fig:LLratios} we show the factor in square brackets in~\cref{eq:CLL} normalised by its non-interacting value.
Deviations from unity therefore quantify the effect of the measured final-state interactions for the conversion from finite to infinite volume at each energy level.
Note that the lowest energy levels for the $\vP^2=0,4$ frames lie very close to the two-particle threshold where the pseudo-phase varies rapidly, making this ratio highly sensitive to the derivative of the scattering phase. 
This produces near-cancellations in the ratio such that these values are too small to be seen in~\cref{fig:LLratios}.

\begin{table}[th]
\centering
\begin{tabular}{c|cc|cc|cc}
\hline
& \multicolumn{2}{c|}{\tt a12m412} & \multicolumn{2}{c|}{\tt a094m412} & \multicolumn{2}{c}{\tt a064m412} \\
State & $q$ & $\mathcal{C}_{LL}$ & $q$ & $\mathcal{C}_{LL}$ & $q$ & $\mathcal{C}_{LL}$ \\
\hline
$\vP^2=0$, $n=0$ & $0.071(31)$ & $62(59)$ & $0.054(46)$ & $146(119)$ & $0.117(16)$ & $109(15)$ \\
$\vP^2=0$, $n=1$ & $1.019(11)$ & $73.95(58)$ & $1.0415(57)$ & $78.20(29)$ & $1.0472(36)$ & $80.01(16)$ \\
$\vP^2=0$, $n=2$ & $1.454(32)$ & $63.40(73)$ & $1.4766(76)$ & $65.79(13)$ & $1.4755(57)$ & $66.84(11)$ \\
$\vP^2=0$, $n=3$ & $1.746(29)$ & $88.5(55)$ & $1.7757(94)$ & $83.3(24)$ & $1.7840(86)$ & $82.5(21)$ \\
$\vP^2=0$, $n=4$ & $2.057(31)$ & $97(19)$ & $2.062(60)$ & $96(22)$ & $2.008(11)$ & $130.1(80)$ \\
\hline
$\vP^2=1$, $n=0$ & $0.440(21)$ & $90.1(36)$ & $0.472(10)$ & $91.2(15)$ & $0.4896(77)$ & $91.6(10)$ \\
$\vP^2=1$, $n=1$ & $1.160(12)$ & $67.74(95)$ & $1.1571(40)$ & $70.16(32)$ & $1.1688(65)$ & $72.45(57)$ \\
$\vP^2=1$, $n=2$ & $1.432(49)$ & $107(71)$ & $1.4409(75)$ & $146(77)$ & $1.472(17)$ & $117(150)$ \\
$\vP^2=1$, $n=3$ & $1.525(34)$ & $74(28)$ & $1.5379(57)$ & $76.09(16)$ & $1.5455(81)$ & $77.27(14)$ \\
\hline
$\vP^2=2$, $n=0$ & $0.601(12)$ & $91.6(92)$ & $0.6194(76)$ & $109(12)$ & $0.6275(69)$ & $121(12)$ \\
$\vP^2=2$, $n=1$ & $0.710(11)$ & $97(16)$ & $0.7254(32)$ & $88.9(19)$ & $0.7299(49)$ & $88.3(26)$ \\
$\vP^2=2$, $n=2$ & $1.199(12)$ & $159(135)$ & $1.2007(38)$ & $195(75)$ & $1.2033(67)$ & $272(164)$ \\
$\vP^2=2$, $n=3$ & $1.262(12)$ & $75.3(42)$ & $1.2588(34)$ & $77.86(19)$ & $1.2671(79)$ & $79.24(27)$ \\
$\vP^2=2$, $n=4$ & $1.601(61)$ & $93(43)$ & $1.5746(60)$ & $104.8(91)$ & $1.5873(81)$ & $100.7(26)$ \\
\hline
$\vP^2=3$, $n=0$ & $0.712(19)$ & $109(50)$ & $0.7155(77)$ & $111(16)$ & $0.736(14)$ & $167(75)$ \\
$\vP^2=3$, $n=1$ & $0.896(27)$ & $54.4(28)$ & $0.9006(71)$ & $56.70(45)$ & $0.9115(61)$ & $57.27(28)$ \\
$\vP^2=3$, $n=2$ & $1.660(32)$ & $74(90)$ & $1.678(21)$ & $58(97)$ & $1.682(15)$ & $58(38)$ \\
\hline
$\vP^2=4$, $n=0$ & $0.128(52)$ & $70(42)$ & $0.077(37)$ & $75(356)$ & $0.092(56)$ & $43(695)$ \\
$\vP^2=4$, $n=1$ & $0.801(56)$ & $101(27)$ & $0.774(32)$ & $94(18)$ & $0.818(11)$ & $115.7(80)$ \\
$\vP^2=4$, $n=2$ & $1.020(13)$ & $82.1(20)$ & $1.0125(94)$ & $84.8(10)$ & $1.0301(53)$ & $88.58(85)$ \\
$\vP^2=4$, $n=3$ & $1.450(15)$ & $75.1(36)$ & $1.4275(87)$ & $88(14)$ & $1.443(11)$ & $80.0(34)$ \\
\hline
\end{tabular}
\caption{Relative momenta and Lellouch--L\"uscher factors on the three ensembles for all finite-volume energy levels  for $I=3/2$ SU(3)$_{\rm F}$ $K\pi$ scattering using the SL phase-shift parametrisation.}
\label{tab:LLfactors}
\end{table}

\subsection{Decay amplitude extraction}
\label{sec:amplitude}

\begin{figure}[th]
    \centering
    \includegraphics[width=0.42\textwidth]{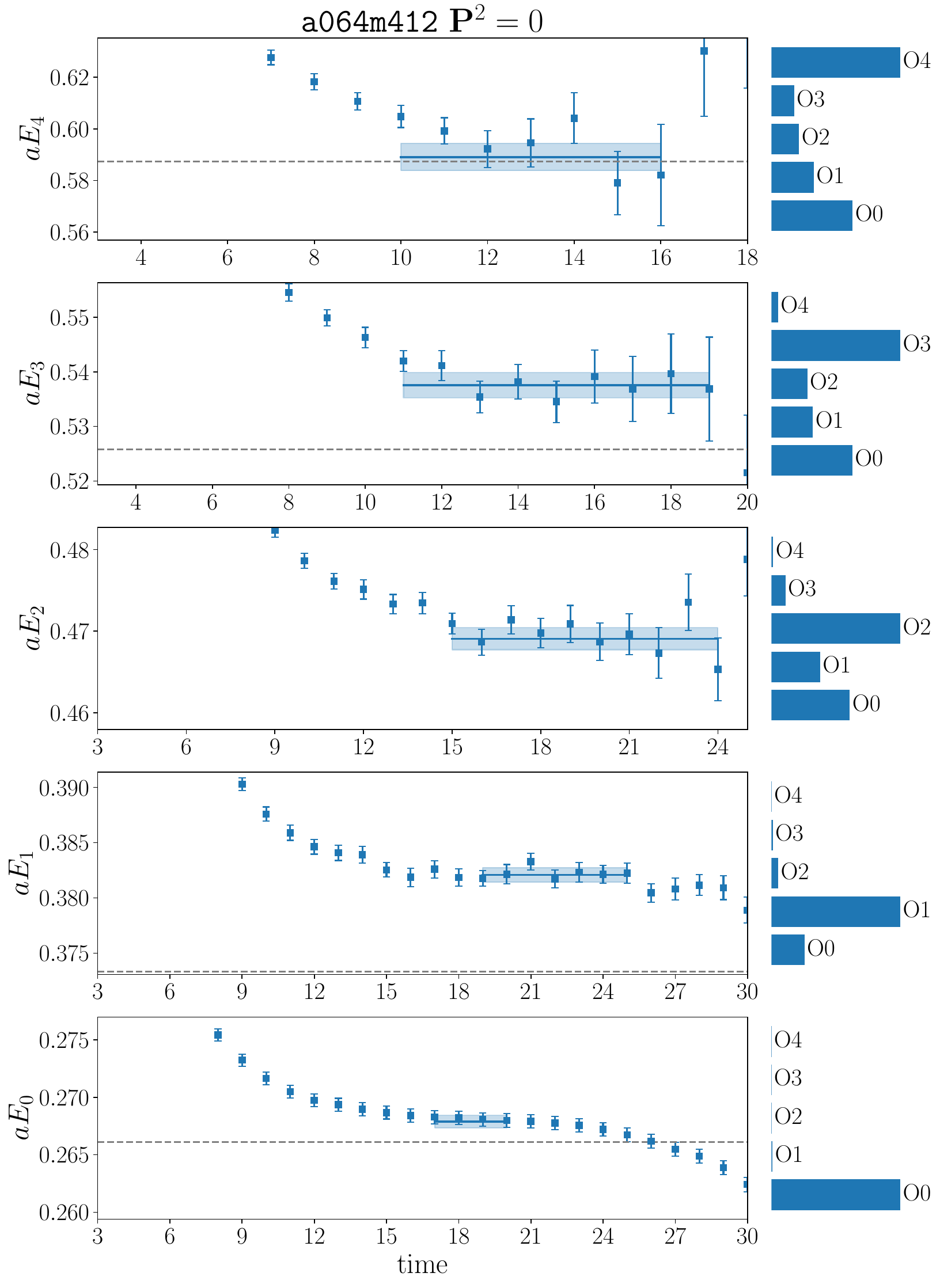}
    \includegraphics[width=0.42\textwidth]{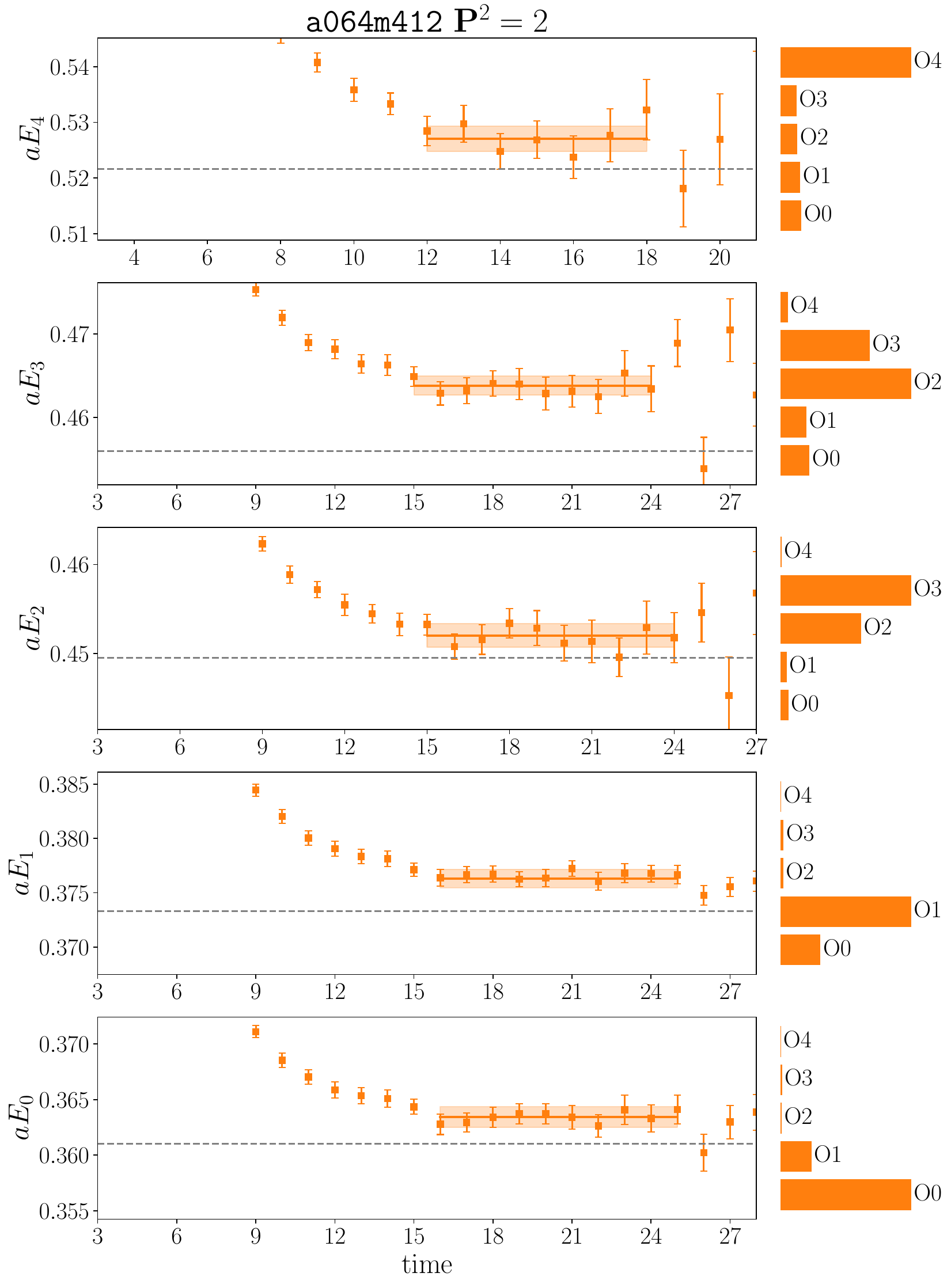}
    \caption{Effective energies of the GEVP eigenvalues $\lambda_n^\vP(t)$ and the relative contributions from each interpolating operator on the {\tt a064m412} ensemble for the $\vP^2=0$ (left) and $\vP^2=2$ (right) frames.}
    \label{fig:GEVP_hist}
\end{figure}

The finite-volume matrix element entering the Lellouch--L\"uscher relation is obtained by projecting the three-point correlator onto the individual $K\pi$ finite-volume eigenstates.
To this end, we use the eigenvectors determined from the GEVP of the two-point correlation matrix in~\cref{sec:kpi_spectrum} to construct optimised sink operators,
\begin{align}
    \widetilde O_n^{\vP}(t) &= \sum_{I=1}^{n_{\rm op}} \left[u_n^{\vP}(t^\star,t_0)\right]^*_I O_I^{\vP}(t),  \label{eq:projected_op} \\
    \widetilde C^{\vP,Q_i}_{3,n}(t_{K\pi},t_H,t_D) &= \sum_{I=1}^{n_{\rm op}} \left[u_n^{\vP}(t^\star,t_0)\right]^*_I C^{\vP,Q_i}_{3,I}(t_{K\pi},t_H,t_D), \label{eq:projected_3pt}
\end{align}
where $C^{\vP}_{3,I}$ denotes the three-point correlator with the $Q_i$ four-quark operator at time $t_D<t_H<t_{K\pi}$ and the $I^{\rm th}$ $K\pi$ interpolator at the sink. 
The GEVP eigenvectors are evaluated at a fixed GEVP reference time $t_0$ and diagonalisation time $t^\star$, which is chosen to yield precise and accurate overlap with the particular energy level.
The resulting linear combination has maximal overlap with the state of energy $E_n^\vP$, while suppressing contributions from the remaining states in the operator basis.
The relative weights of the initial $K\pi$ interpolators to the optimal basis are shown for the example of the $\vP^2=0$ and $\vP^2=2$ frames on the {\tt a064m412} ensemble in~\cref{fig:GEVP_hist}.

The optimally-projected correlator $\widetilde C^{\vP,Q_i}_{3,n}$ isolates the transition $D\to(K\pi)_n$ for the $n^{\rm th}$ $K\pi$ energy level, from which the corresponding finite-volume matrix element for this transition can be extracted.
The spectral form of this correlator is
\begin{equation}
    \widetilde C^{\vP,Q_i}_{3,n}(t_{K\pi},t_H,t_D) =  \frac{Z_n^{\vP}\,Z_D^{\vP *}}{4E_n^{\vP}E_D^{\vP}}\,\langle E_n,\vP,L|Q_i|D,\vP\rangle\, e^{-E_n^{\vP} (t_{K\pi}-t_{H})} e^{-E_D^{\vP}(t_H-t_D)} +\cdots ,
\end{equation}
where we have omitted higher-energy contributions of both the $D$ and $K\pi$ states.
Given a sufficient separation between $t_D$ and $t_{K\pi}$, this ground state will dominate the signal.
To isolate the ground-state dominant behaviour and cancel the Euclidean-time dependence, we can form a ratio
\begin{equation}
    R_n^{\vP,Q_i}(t_{K\pi},t_H,t_D) = Z_n^{\vP}\,Z_D^{\vP *}\frac{\widetilde C^{\vP,Q_i}_{3,n}(t_{K\pi},t_H,t_D)}{C_{K\pi,n}^{\vP}(t_{K\pi},t_H)\,C_D^{\vP}(t_H,t_D)} \to \langle E_n,\vP,L|Q_i|D,\vP\rangle,
\end{equation}
where the arrow indicates the limits $t_D\ll t_H\ll t_{K\pi}$ and $t_{K\pi}-t_D\to\infty$, $C_{K\pi,n}^{\vP}$ is the optimally-projected $K\pi$ correlator, $C_D^\vP$ is the $D$-meson correlator, and $Z_n^\vP$ and $Z_D^\vP$ are their respective overlap factors extracted from fitting these two-point functions.
Note that due to the projection of both the three-point and $K\pi$ two-point functions, there is an implicit dependence on $t_0$.

Work to produce and analyse these three-point functions and their ratios is ongoing, which will result in a set of finite-volume transition amplitudes $\langle E_n,\vP,L|Q_i|D,\vP\rangle$ for each energy level $n$ and operator $Q_i$. 
Using the Lellouch--L\"uscher factors shown in~\cref{tab:LLfactors}, these can be converted to the infinite volume.
We plan to renormalise these matrix elements using the RI-MOM and RI-SMOM schemes before extrapolating our results to the continuum and the physical decay point at the $D$ meson mass.
From the modest dataset for this current calculation, we anticipate final relative uncertainties to be $O(10\%)$.
Future calculations with more gauge-field configurations and more lattice ensembles can improve this precision further.

\vspace{-10pt}
\section{Conclusions and outlook}
\label{sec:conclusions}

In this work, we have determined the elastic $(K\pi)_{\mathbf{27}}$ scattering amplitude at the \SUF-symmetric point, described how the same finite-volume spectrum and corresponding operators enter the calculation of the $D\to(K\pi)_{\mathbf{27}}$ decay amplitude, and outlined various technical details of our method for determining the latter. In a forthcoming manuscript we will complete the calculation to give a first systematic prediction of the decay amplitude at the flavour-symmetric point.

The two key numerical results presented in this work are the continuum scattering length $a_0=0.926(59)_{\rm stat}(53)_{\rm sys}\,{\rm GeV}^{-1}$, and the corresponding scattering phase extrapolated to the $D$-meson mass $\delta_{\mathbf{27}}(M_D)=-38.4(2.4)^\circ$.
Given the extensive published literature concerning lattice QCD scattering calculations, the main novelty of this work is our application of two continuum-limit strategies: extrapolating the scattering parameters directly and extrapolating the finite-volume spectrum before applying the scattering formalism. The two approaches give compatible results and the scattering-length-only parametrisation gives a good description of our continuum extrapolated energies.

The use of five different total momenta in the finite-volume frame led to a total of 21 continuum extrapolated finite-volume energies.
This gives strong coverage of the scattering phase from threshold up to approximately $4M_\pi$, constrains the energy dependence of the phase more reliably in the elastic regime, and thereby improves the extrapolation required to determine its value at the $D$-meson mass. At the same time it should be emphasised that the four-pion threshold is not included in this step. The threshold is known to introduce a cusp in the scattering phase, with the discontinuity in the derivative set by the two-to-four interaction strength. We expect the effect of this to be subdominant to other sources of uncertainty in the present dataset, and leave a careful consideration of its effect to future work~\cite{Mukherjee:2026fourpion}.
The scattering analysis provides the finite-volume input required for the decay calculation. In particular, the phase-shift determines the Lellouch--L\"uscher factors needed to convert the finite-volume weak matrix elements into the infinite-volume amplitude.

As discussed in~\cref{sec:dkpi_decay} for the three-point calculation, our distillation construction accommodates local four-quark insertions without requiring the full four-quark fields to be stored. The $\overline{\mathbf{15}}\to\mathbf{27}$ transition is the most straightforward as it avoids self-contractions and the associated power-divergent subtractions as well as additional resonant behaviour. 
A subsequent manuscript will complete the calculation by analysing the three-point functions obtained using the computational strategy of~\cref{sec:dkpi_decay}, renormalising $Q_1$ and $Q_2$, applying the Lellouch--L\"uscher factors listed in~\cref{tab:LLfactors}, and finally taking the continuum limit of the $D\to(K\pi)_{\mathbf{27}}$ decay amplitude.


\section*{Acknowledgments}

We thank
Peter Boyle,
Matteo Di Carlo,
Ryan Hill,
Raoul Hodgson, and
Teseo San Jose
for useful discussions and the OpenLat Initiative for making their gauge-field ensembles available to us.
F.E.\, has received funding from the European Union's Horizon Europe research and innovation programme under the Marie Sk\l{}odowska-Curie grant agreement No.~101106913.
A.P.\, and F.E.\, received funding from the European Research Council (ERC) under the European Union’s Horizon 2020 research and innovation program under Grant Agreement No. 757646. N.P.L.\, and A.P.\, received funding from the European Research Council (ERC) under the European Union’s Horizon 2020 research and innovation program under Grant Agreement No. 813942.
M.B., M.T.H., and A.P\, are supported by UK STFC grant ST/X000494/1. 
M.T.H.\ and R.M.\ are supported by UKRI Future Leaders Fellowship MR/T019956/1.
S.P is partially supported from the projects IMAGE-N (EXCELLENCE/0524/0459), StrongILA and partonWF (VISION ERC/0525/0010) co-financed by the European Regional Development Fund and the Republic of Cyprus through the Research and Innovation Foundation within the framework of the Cohesion Policy Programme “THALIA 2021-2027”. S.P.\ acknowledges support by DOE
Grant KA2401045, when he was at University of Maryland.
N.P.L.\ acknowledges support from the U.K. Science and Technology Facilities Council (STFC) grant numbers ST/T000694/1, ST/X000664/1.

This work was supported by a grant from the Swiss National Supercomputing Centre (CSCS) under project ID lp141 on Alps as well as the DiRAC Extreme Scaling services (Tesseract \& Tursa) at the University of Edinburgh, managed by the EPCC on behalf of the STFC DiRAC HPC Facility (\url{www.dirac.ac.uk}). 
The DiRAC service at Edinburgh was funded by BEIS, UKRI and STFC capital funding and STFC operations grants. DiRAC is part of the UKRI Digital Research Infrastructure.
This work was performed in part using the flagship cluster Zaratan and the storage resources of the University of Maryland High Performance Computing Cluster (HPCC), which is sustained by the Division of Information Technology and the University.

\clearpage

\appendix


\section{GEVP fit results}
\label{app:GEVPtables}

\begin{figure}[th]
    \centering
    \includegraphics[width=0.32\textwidth]{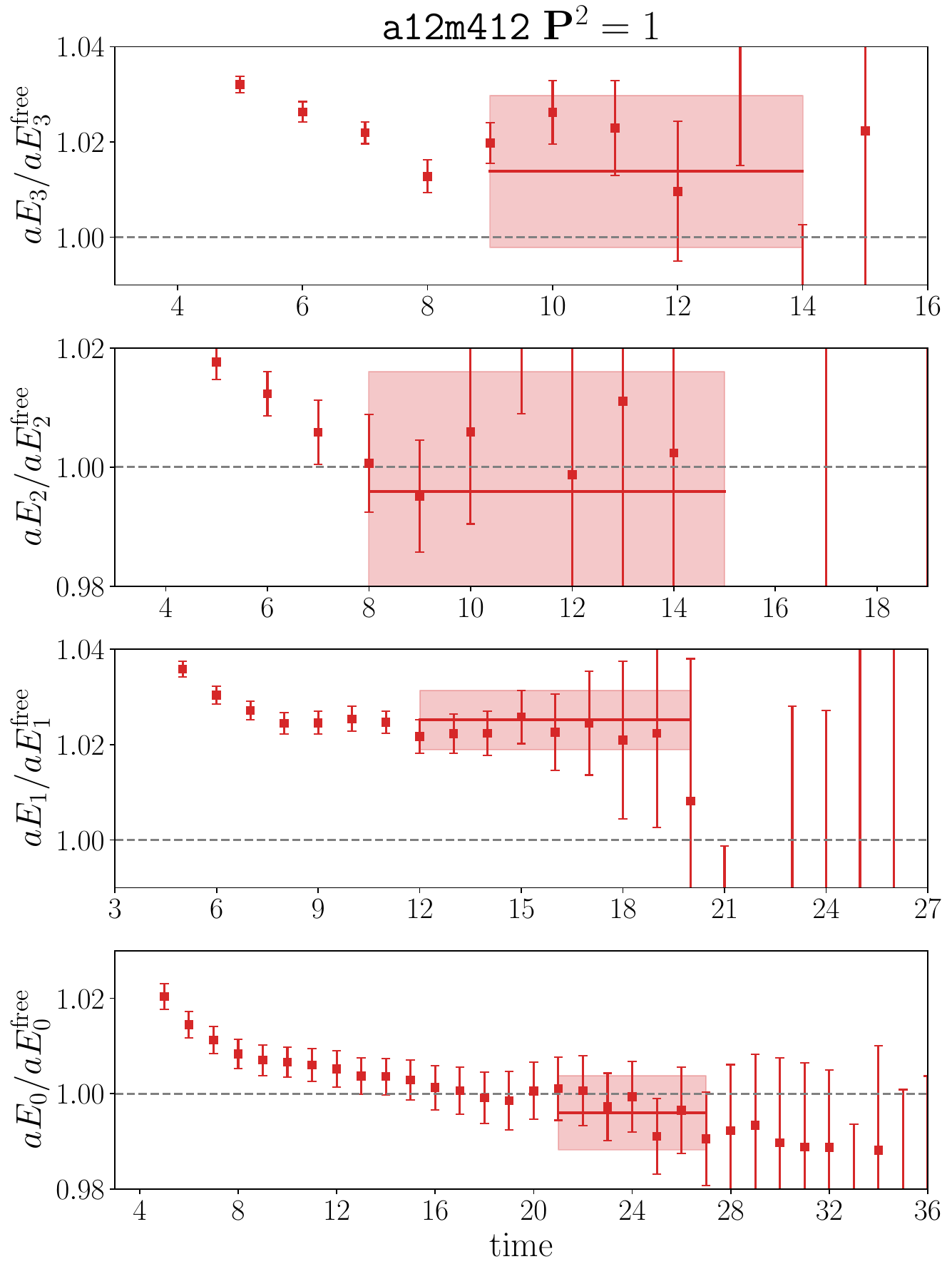}
    \includegraphics[width=0.32\textwidth]{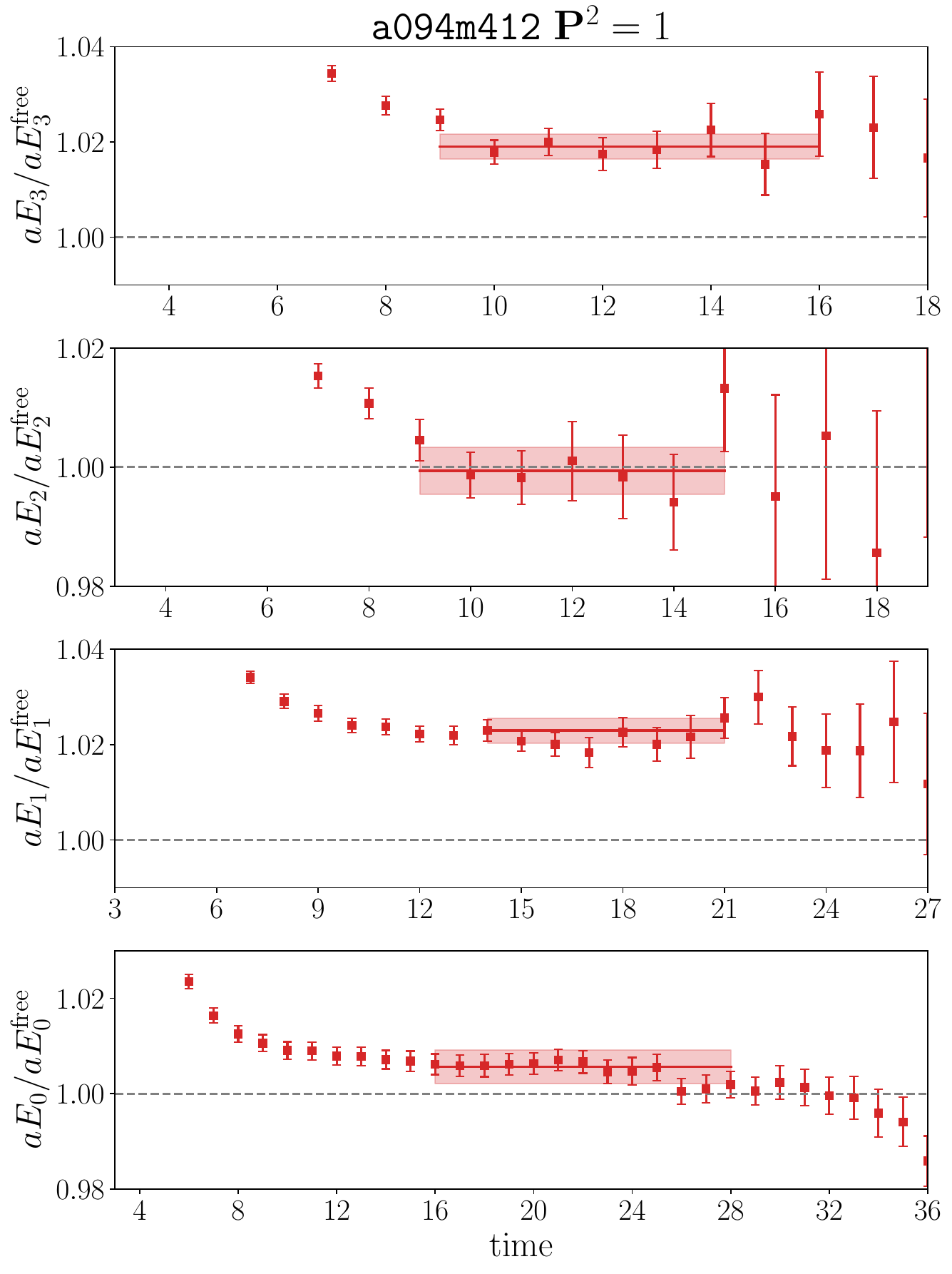}
    \includegraphics[width=0.32\textwidth]{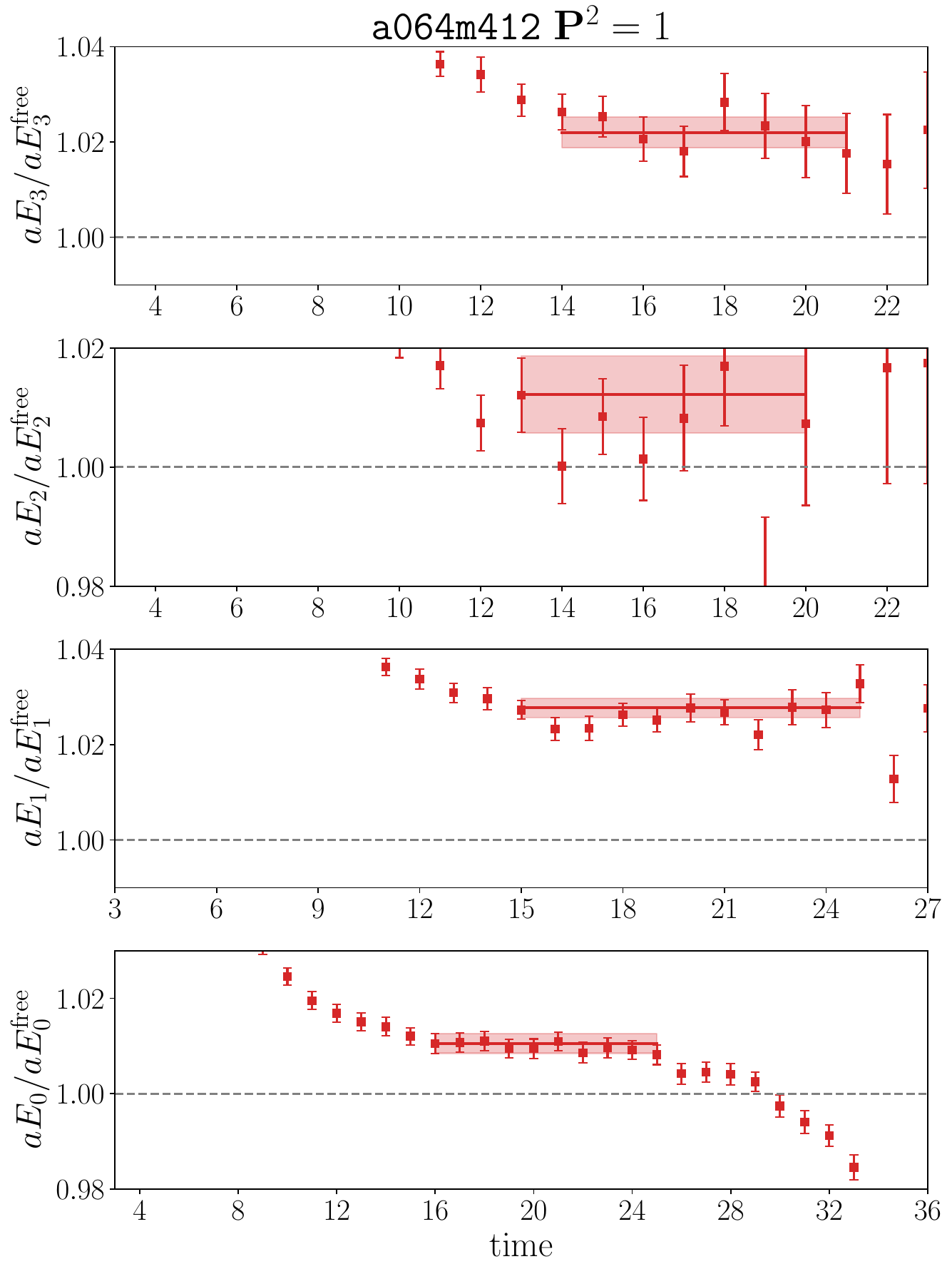}
    \caption{Effective energies of the GEVP eigenvalues $\lambda_n^\vP(t)$ and their fitted finite-volume energy levels normalised to the non-interacting energy levels indicated by the dashed lines for the $\vP^2=1$ frame. The {\tt a12m412} ensemble is shown in the left column, the {\tt a094m412} in the middle, and {\tt a064m412} on the right. }
    \label{fig:GEVP_nonintRAT_P1}
\end{figure}

\begin{figure}[th]
    \centering
    \includegraphics[width=0.32\textwidth]{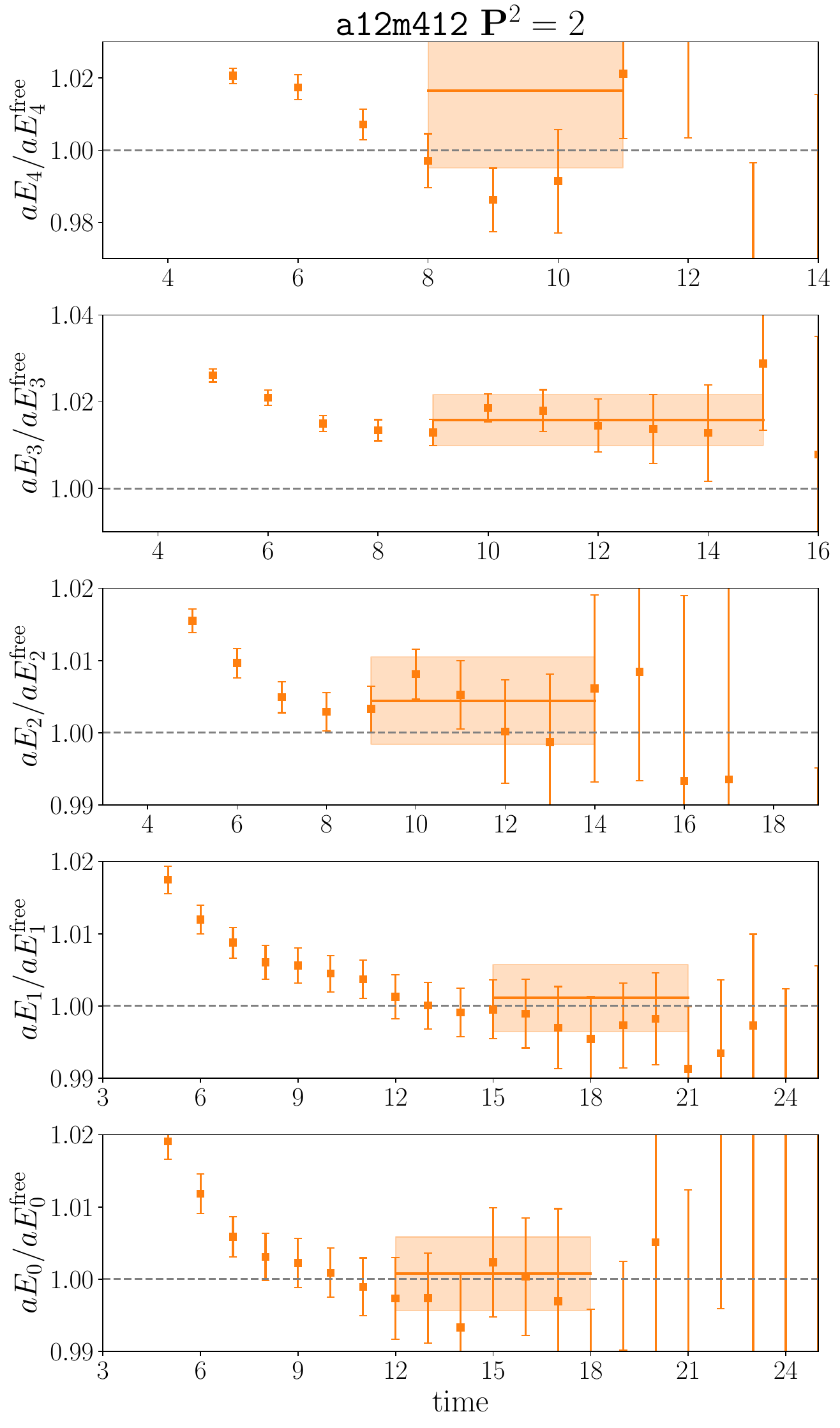}
    \includegraphics[width=0.32\textwidth]{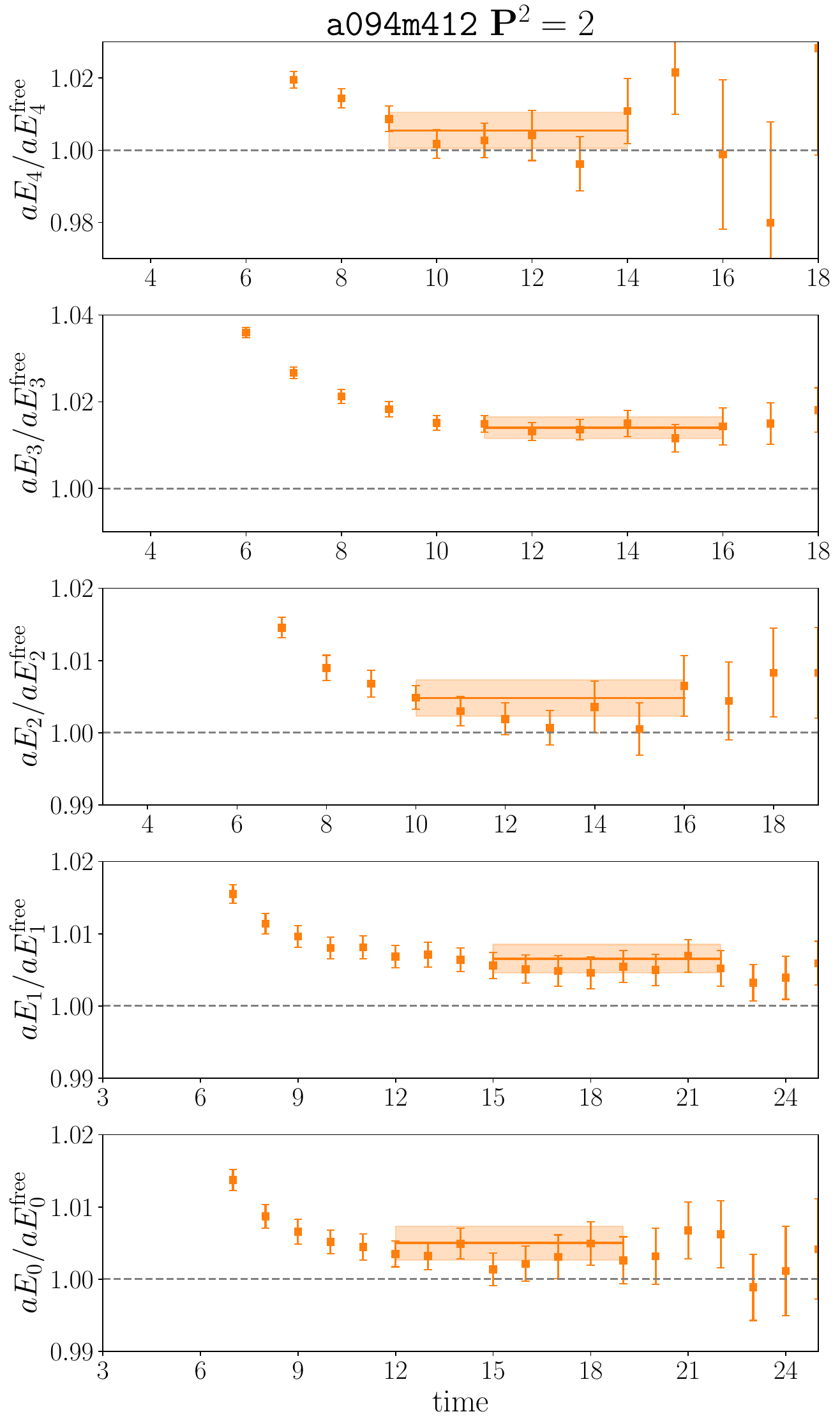}
    \includegraphics[width=0.32\textwidth]{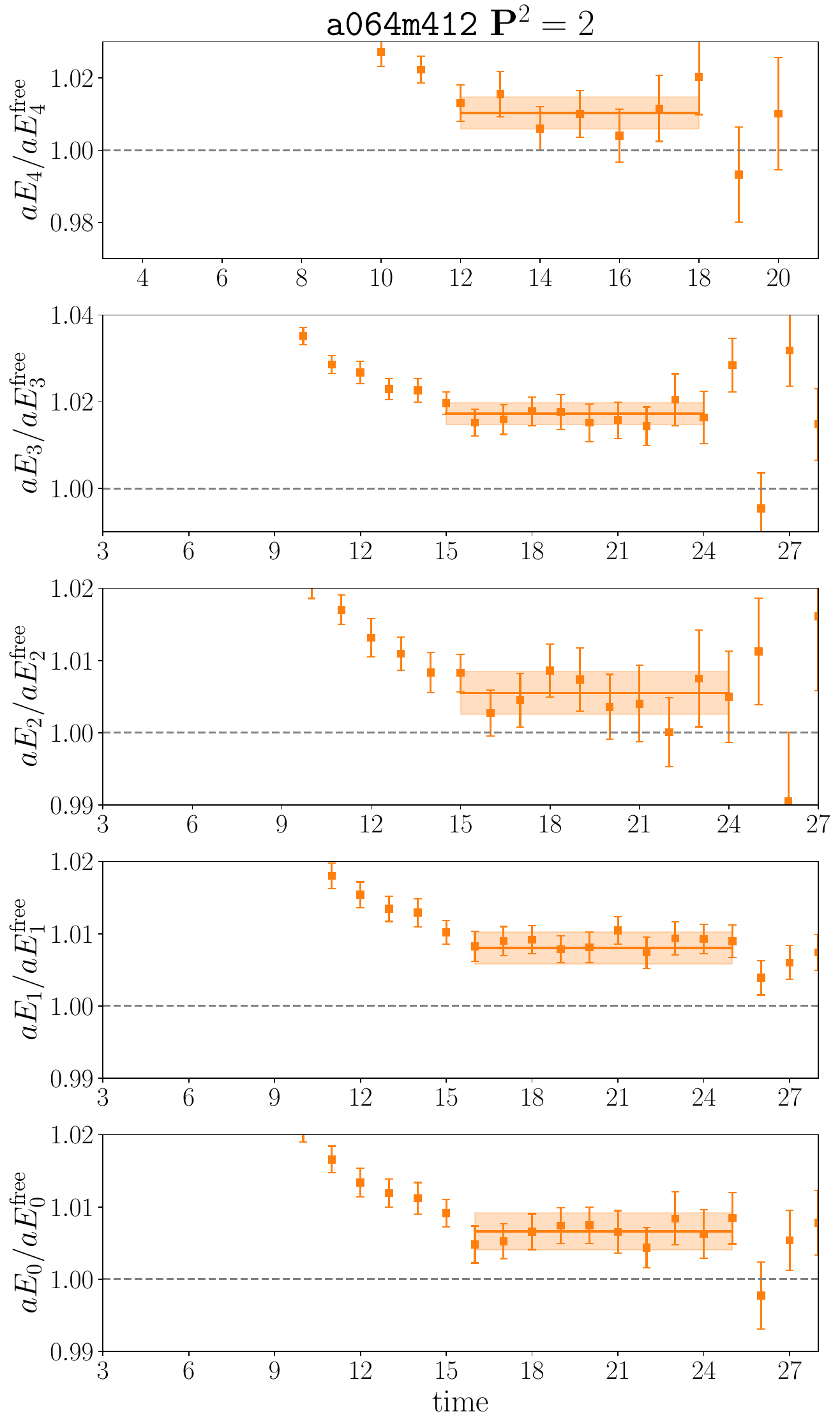}
    \caption{Effective energies of the GEVP eigenvalues $\lambda_n^\vP(t)$ and their fitted finite-volume energy levels normalised to the non-interacting energy levels indicated by the dashed lines for the $\vP^2=2$ frame. The {\tt a12m412} ensemble is shown in the left column, the {\tt a094m412} in the middle, and {\tt a064m412} on the right.}
    \label{fig:GEVP_nonintRAT_P2}
\end{figure}

\begin{figure}[th]
    \centering
    \includegraphics[width=0.32\textwidth]{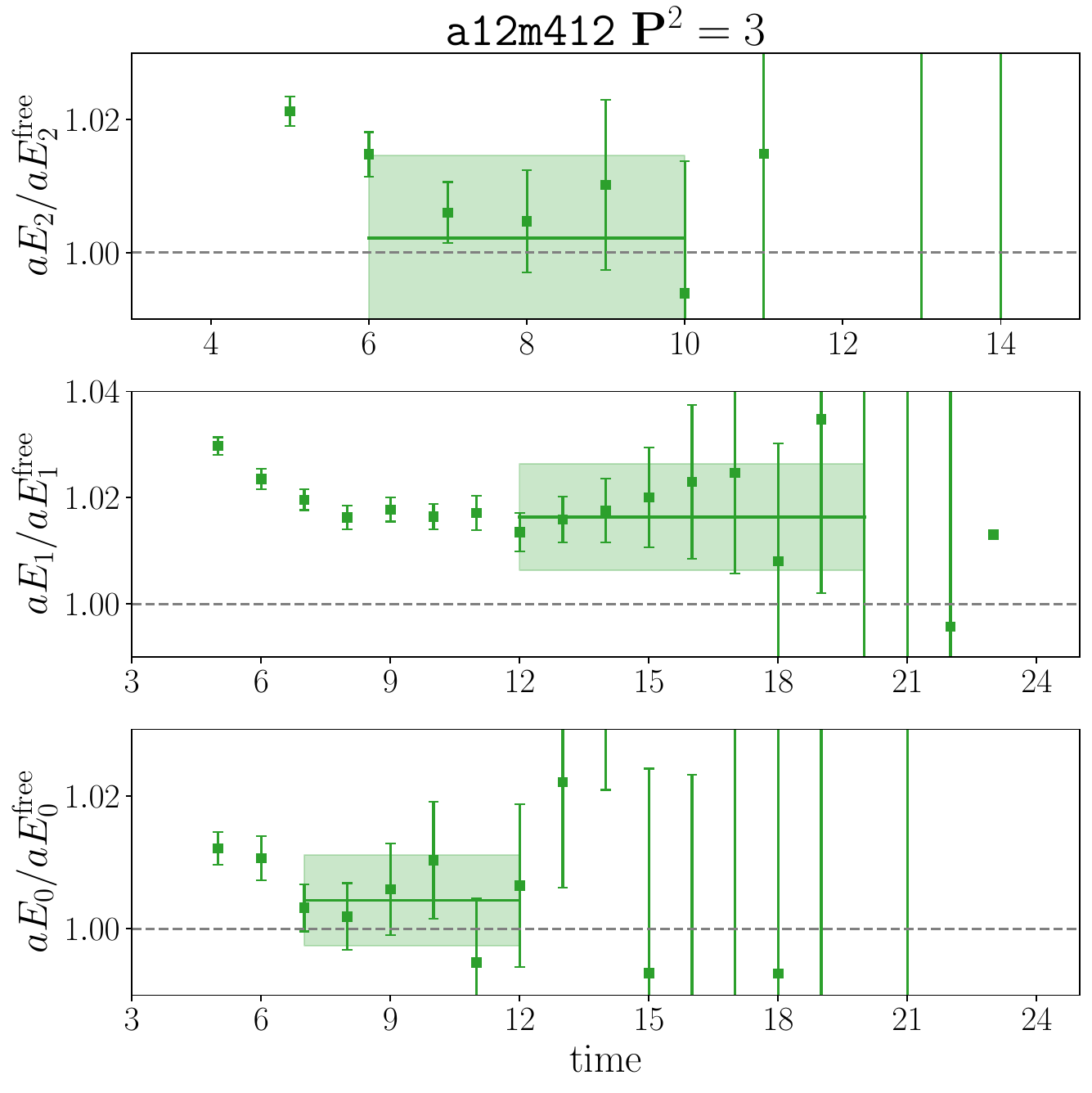}
    \includegraphics[width=0.32\textwidth]{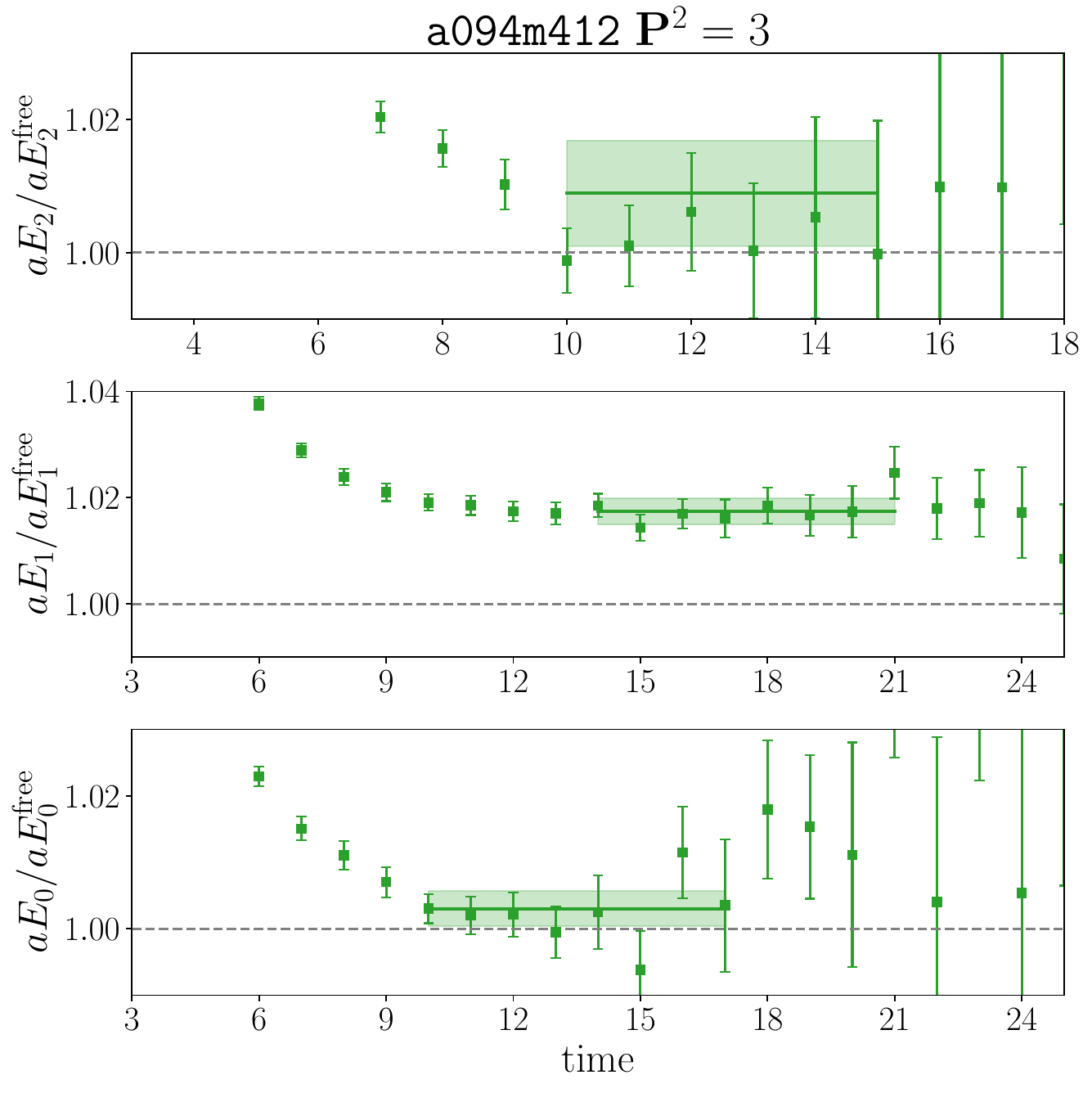}
    \includegraphics[width=0.32\textwidth]{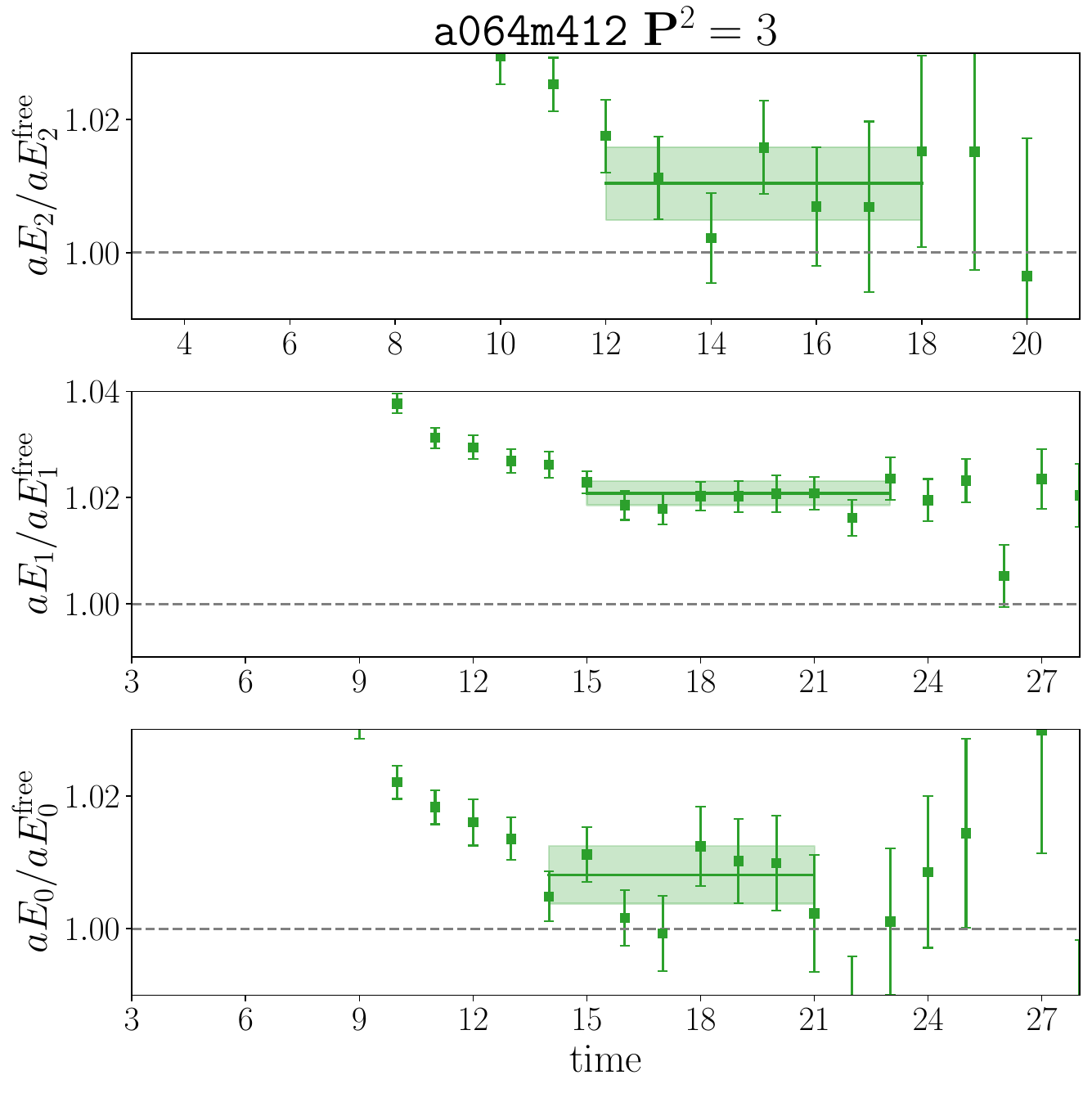}
    \caption{Effective energies of the GEVP eigenvalues $\lambda_n^\vP(t)$ and their fitted finite-volume energy levels normalised to the non-interacting energy levels indicated by the dashed lines for the $\vP^2=3$ frame. The {\tt a12m412} ensemble is shown in the left column, the {\tt a094m412} in the middle, and {\tt a064m412} on the right.}
    \label{fig:GEVP_nonintRAT_P3}
\end{figure}

\begin{figure}[th]
    \centering
    \includegraphics[width=0.32\textwidth]{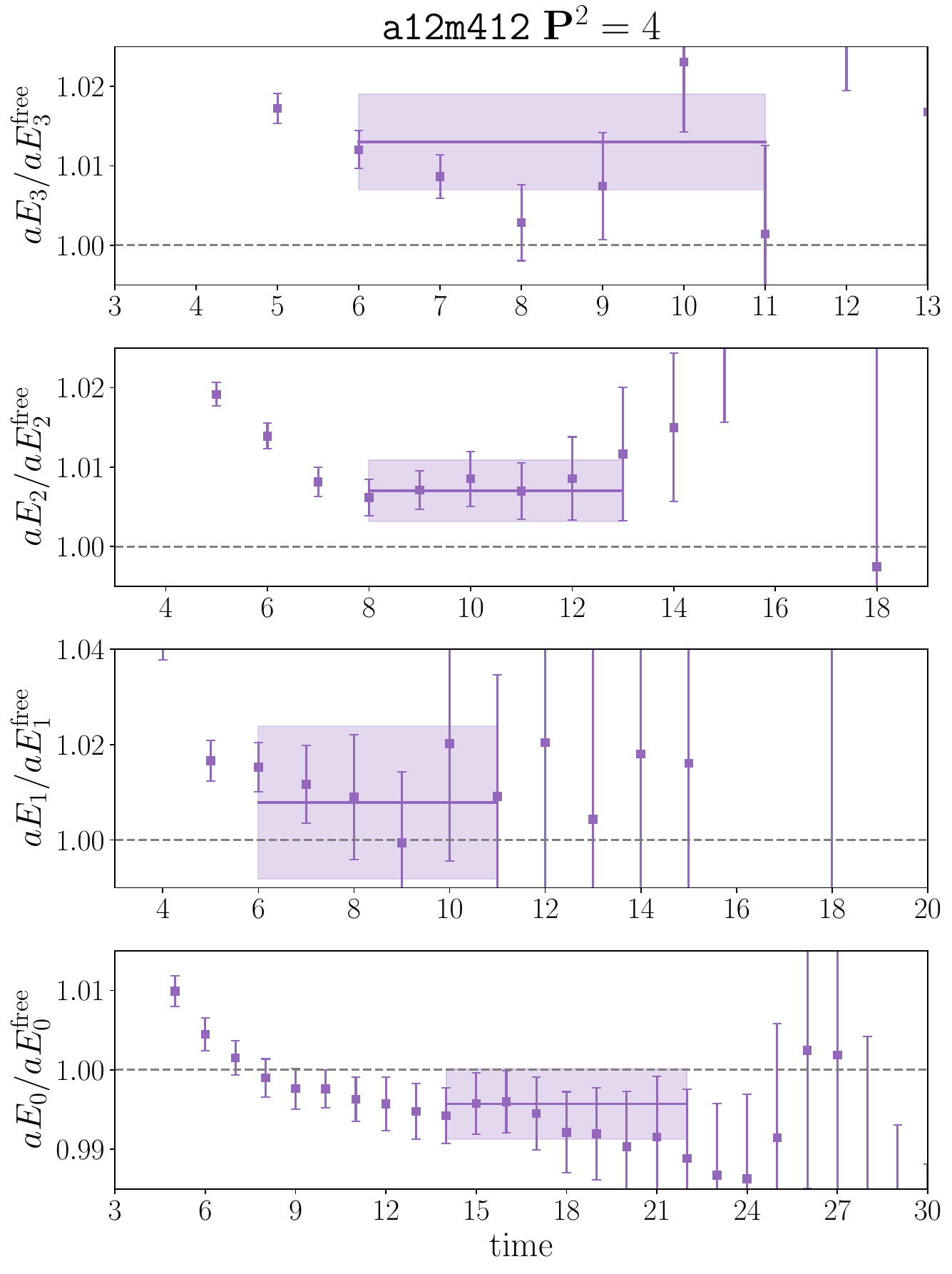}
    \includegraphics[width=0.32\textwidth]{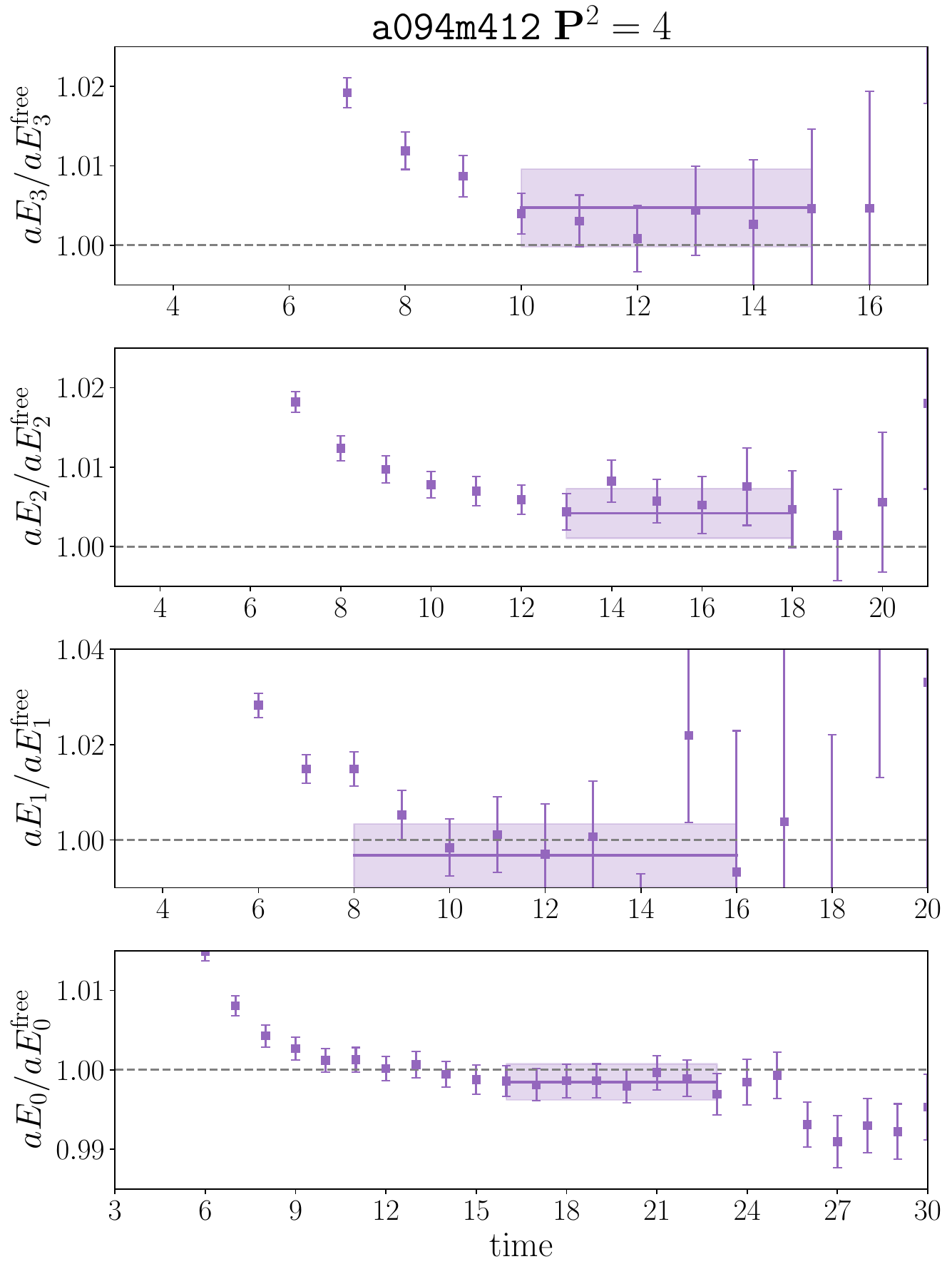}
    \includegraphics[width=0.32\textwidth]{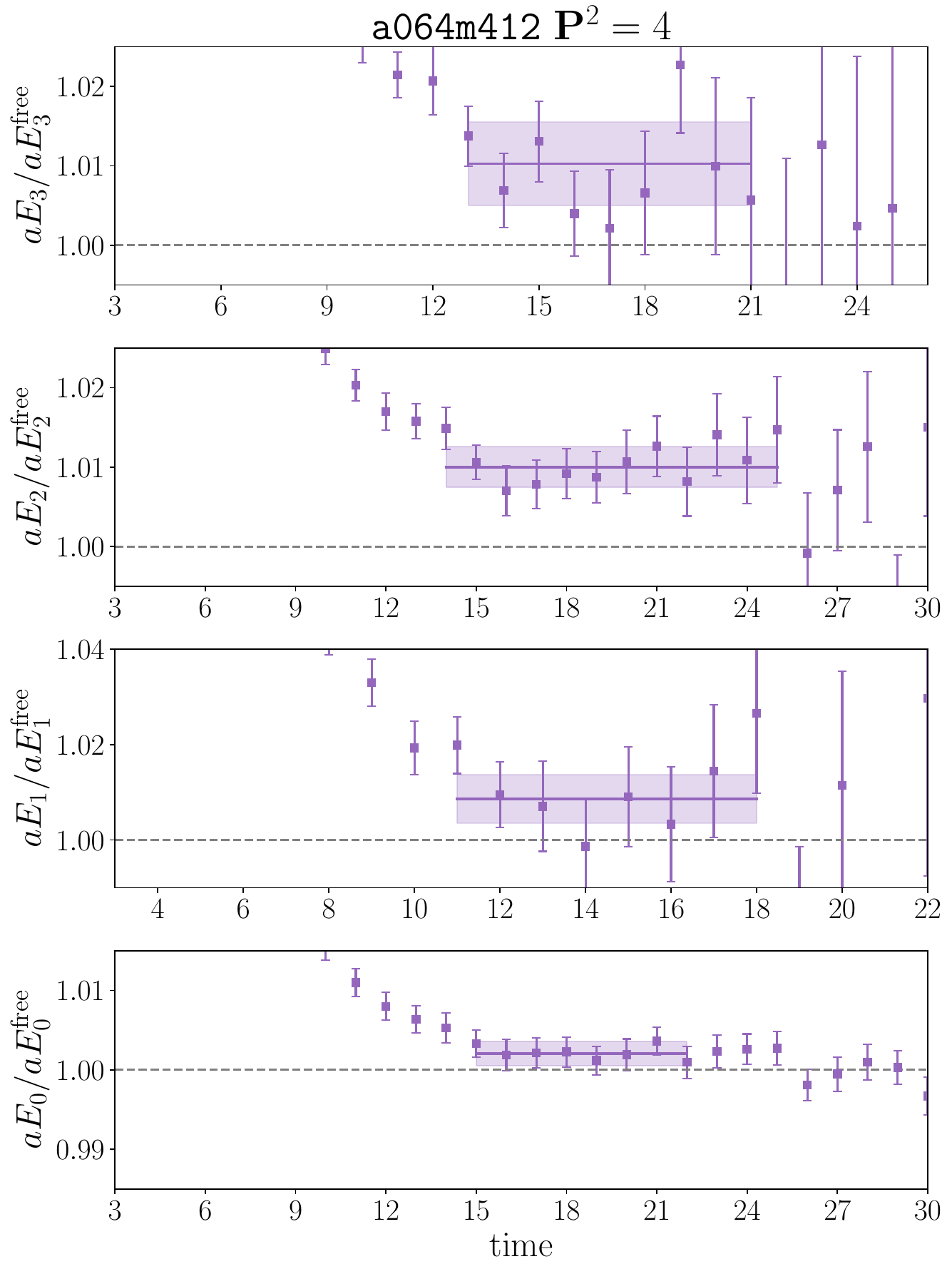}
    \caption{Effective energies of the GEVP eigenvalues $\lambda_n^\vP(t)$ and their fitted finite-volume energy levels normalised to the non-interacting energy levels indicated by the dashed lines for the $\vP^2=4$ frame. The {\tt a12m412} ensemble is shown in the left column, the {\tt a094m412} in the middle, and {\tt a064m412} on the right. }
    \label{fig:GEVP_nonintRAT_P4}
\end{figure}

\begin{table}[th]
    \centering
    \begin{tabular}{ccccc}
        \hline
        State & $aE_{n}^{\vP}(L)$ & $[t_{\min},t_{\max}]$ & Best $\chi^2_{\rm /dof}$ & Best $p$-value \\
        \hline
        $\vP^2=0$, $n=0$ & $0.5015(21)$ & $[24,31]$ & 1.016 [16] & 0.435 \\
        $\vP^2=0$, $n=1$ & $0.7312(41)$ & $[15,21]$ &  &  \\
        $\vP^2=0$, $n=2$ & $0.911(15)$ & $[11,16]$ &  &  \\
        $\vP^2=0$, $n=3$ & $1.042(15)$ & $[7,13]$ &  &  \\
        $\vP^2=0$, $n=4$ & $1.188(16)$ & $[5,10]$ &  &  \\
        \hline
        $\vP^2=1$, $n=0$ & $0.6097(47)$ & $[21,28]$ & 1.008 [17] & 0.446 \\
        $\vP^2=1$, $n=1$ & $0.8292(50)$ & $[12,21]$ &  &  \\
        $\vP^2=1$, $n=2$ & $0.938(19)$ & $[8,16]$ &  &  \\
        $\vP^2=1$, $n=3$ & $0.978(15)$ & $[9,15]$ &  &  \\
        \hline
        $\vP^2=2$, $n=0$ & $0.6974(36)$ & $[12,19]$ & 0.371 [18] & 0.993 \\
        $\vP^2=2$, $n=1$ & $0.7249(34)$ & $[15,22]$ &  &  \\
        $\vP^2=2$, $n=2$ & $0.8838(53)$ & $[9,15]$ &  &  \\
        $\vP^2=2$, $n=3$ & $0.9077(53)$ & $[9,16]$ &  &  \\
        $\vP^2=2$, $n=4$ & $1.044(22)$ & $[8,12]$ &  &  \\
        \hline
        $\vP^2=3$, $n=0$ & $0.7712(52)$ & $[7,13]$ & 0.341 [12] & 0.982 \\
        $\vP^2=3$, $n=1$ & $0.8221(81)$ & $[12,21]$ &  &  \\
        $\vP^2=3$, $n=2$ & $1.101(14)$ & $[6,11]$ &  &  \\
        \hline
        $\vP^2=4$, $n=0$ & $0.7210(32)$ & $[14,23]$ & 0.544 [17] & 0.932 \\
        $\vP^2=4$, $n=1$ & $0.837(13)$ & $[6,12]$ &  &  \\
        $\vP^2=4$, $n=2$ & $0.8999(34)$ & $[8,14]$ &  &  \\
        $\vP^2=4$, $n=3$ & $1.0492(62)$ & $[6,12]$ &  &  \\
        \hline
    \end{tabular}
    \caption{Fit results for the centre-of-mass finite-volume energy levels extracted from the GEVP using a combined model-averaging procedure for each moving frame on the {\tt a12m412} ensemble. $[t_{\min},t_{\max}]$ indicate the minimum and maximum timeslices entering the set of fits performed. The best $\chi^2_{\rm /dof}$ and associated $p$-value are also shown.}
    \label{tab:gevp-energies-a12}
\end{table}

\begin{table}[th]
    \centering
    \begin{tabular}{ccccc}
        \hline
        State & $aE_{n}^{\vP}(L)$ & $[t_{\min},t_{\max}]$ & Best $\chi^2_{\rm /dof}$ & Best $p$-value \\
        \hline
        $\vP^2=0$, $n=0$ & $0.3908(12)$ & $[17,27]$ & 0.460 [12] & 0.938 \\
        $\vP^2=0$, $n=1$ & $0.5653(15)$ & $[15,27]$ &  &  \\
        $\vP^2=0$, $n=2$ & $0.6990(24)$ & $[10,18]$ &  &  \\
        $\vP^2=0$, $n=3$ & $0.7991(37)$ & $[8,13]$ &  &  \\
        $\vP^2=0$, $n=4$ & $0.899(22)$ & $[9,13]$ &  &  \\
        \hline
        $\vP^2=1$, $n=0$ & $0.4746(17)$ & $[16,29]$ & 0.427 [17] & 0.980 \\
        $\vP^2=1$, $n=1$ & $0.6303(16)$ & $[14,22]$ &  &  \\
        $\vP^2=1$, $n=2$ & $0.7149(28)$ & $[9,16]$ &  &  \\
        $\vP^2=1$, $n=3$ & $0.7454(19)$ & $[9,17]$ &  &  \\
        \hline
        $\vP^2=2$, $n=0$ & $0.5372(13)$ & $[12,20]$ & 0.669 [21] & 0.868 \\
        $\vP^2=2$, $n=1$ & $0.5572(11)$ & $[15,23]$ &  &  \\
        $\vP^2=2$, $n=2$ & $0.6721(17)$ & $[10,17]$ &  &  \\
        $\vP^2=2$, $n=3$ & $0.6883(17)$ & $[11,17]$ &  &  \\
        $\vP^2=2$, $n=4$ & $0.7821(39)$ & $[9,15]$ &  &  \\
        \hline
        $\vP^2=3$, $n=0$ & $0.5890(16)$ & $[10,18]$ & 0.760 [15] & 0.724 \\
        $\vP^2=3$, $n=1$ & $0.6269(15)$ & $[14,22]$ &  &  \\
        $\vP^2=3$, $n=2$ & $0.8380(66)$ & $[10,16]$ &  &  \\
        \hline
        $\vP^2=4$, $n=0$ & $0.5528(12)$ & $[16,24]$ & 0.585 [16] & 0.898 \\
        $\vP^2=4$, $n=1$ & $0.6316(42)$ & $[8,17]$ &  &  \\
        $\vP^2=4$, $n=2$ & $0.6816(21)$ & $[13,19]$ &  &  \\
        $\vP^2=4$, $n=3$ & $0.7879(38)$ & $[10,16]$ &  &  \\
        \hline
    \end{tabular}
    \caption{Fit results for the centre-of-mass finite-volume energy levels extracted from the GEVP using a combined model-averaging procedure for each moving frame on the {\tt a094m412} ensemble. $[t_{\min},t_{\max}]$ indicate the minimum and maximum timeslices entering the set of fits performed. The best $\chi^2_{\rm /dof}$ and associated $p$-value are also shown.}
    \label{tab:gevp-energies-a094}
\end{table}

\begin{table}[th]
    \centering
    \begin{tabular}{ccccc}
        \hline
        State & $aE_{n}^{\vP}(L)$ & $[t_{\min},t_{\max}]$ & Best $\chi^2_{\rm /dof}$ & Best $p$-value \\
        \hline
        $\vP^2=0$, $n=0$ & $0.26790(50)$ & $[17,21]$ & 0.753 [15] & 0.731 \\
        $\vP^2=0$, $n=1$ & $0.38210(70)$ & $[19,29]$ &  &  \\
        $\vP^2=0$, $n=2$ & $0.4691(13)$ & $[15,25]$ &  &  \\
        $\vP^2=0$, $n=3$ & $0.5375(23)$ & $[11,20]$ &  &  \\
        $\vP^2=0$, $n=4$ & $0.5891(53)$ & $[10,17]$ &  &  \\
        \hline
        $\vP^2=1$, $n=0$ & $0.32310(70)$ & $[16,26]$ & 0.890 [15] & 0.575 \\
        $\vP^2=1$, $n=1$ & $0.42610(80)$ & $[15,26]$ &  &  \\
        $\vP^2=1$, $n=2$ & $0.4862(31)$ & $[13,21]$ &  &  \\
        $\vP^2=1$, $n=3$ & $0.5017(16)$ & $[14,22]$ &  &  \\
        \hline
        $\vP^2=2$, $n=0$ & $0.36340(90)$ & $[16,26]$ & 0.789 [13] & 0.673 \\
        $\vP^2=2$, $n=1$ & $0.37630(80)$ & $[16,26]$ &  &  \\
        $\vP^2=2$, $n=2$ & $0.4520(13)$ & $[15,25]$ &  &  \\
        $\vP^2=2$, $n=3$ & $0.4638(11)$ & $[15,25]$ &  &  \\
        $\vP^2=2$, $n=4$ & $0.5270(23)$ & $[12,19]$ &  &  \\
        \hline
        $\vP^2=3$, $n=0$ & $0.3992(17)$ & $[14,22]$ & 0.129 [10] & 0.999 \\
        $\vP^2=3$, $n=1$ & $0.42330(90)$ & $[15,24]$ &  &  \\
        $\vP^2=3$, $n=2$ & $0.5623(31)$ & $[12,19]$ &  &  \\
        \hline
        $\vP^2=4$, $n=0$ & $0.37410(60)$ & $[15,23]$ & 0.728 [18] & 0.785 \\
        $\vP^2=4$, $n=1$ & $0.4304(22)$ & $[11,19]$ &  &  \\
        $\vP^2=4$, $n=2$ & $0.4605(12)$ & $[14,26]$ &  &  \\
        $\vP^2=4$, $n=3$ & $0.5312(28)$ & $[13,22]$ &  &  \\
        \hline
    \end{tabular}
    \caption{Fit results for the centre-of-mass finite-volume energy levels extracted from the GEVP using a combined model-averaging procedure for each moving frame on the {\tt a064m412} ensemble. $[t_{\min},t_{\max}]$ indicate the minimum and maximum timeslices entering the set of fits performed. The best $\chi^2_{\rm /dof}$ and associated $p$-value are also shown.}
    \label{tab:gevp-energies-a064}
\end{table}

\clearpage
\section{Finite-volume energy continuum limits}
\label{app:CLtables}

\begin{table}[th]
    \centering
    \begin{tabular}{c|ccc|ccc}
        \hline
        & \multicolumn{3}{c|}{$\mathcal{O}(a^2)$ extrapolation} & \multicolumn{3}{c}{$\mathcal{O}(a)$ extrapolation} \\
        State & $E$ [GeV] & $\chi^2_{\rm /dof}$ & $p$-value & $E$ [GeV] & $\chi^2_{\rm /dof}$ & $p$-value \\
        \hline
        $\vP^2=0$, $n=0$ & $0.8288(23)$ & 0.568 [1] & 0.451 & $0.8326(46)$ & 0.420 [1] & 0.517 \\
        $\vP^2=0$, $n=1$ & $1.1864(37)$ & 0.987 [1] & 0.320 & $1.1960(74)$ & 1.481 [1] & 0.224 \\
        $\vP^2=0$, $n=2$ & $1.4509(78)$ & 0.440 [1] & 0.507 & $1.453(15)$ & 0.475 [1] & 0.491 \\
        $\vP^2=0$, $n=3$ & $1.669(11)$ & 0.291 [1] & 0.590 & $1.680(21)$ & 0.420 [1] & 0.517 \\
        $\vP^2=0$, $n=4$ & $1.804(15)$ & 0.242 [1] & 0.623 & $1.773(32)$ & 0.180 [1] & 0.671 \\
        \hline
        $\vP^2=1$, $n=0$ & $0.9187(49)$ & 0.162 [1] & 0.687 & $0.9291(91)$ & 0.381 [1] & 0.537 \\
        $\vP^2=1$, $n=1$ & $1.2547(64)$ & 0.965 [1] & 0.326 & $1.263(12)$ & 0.678 [1] & 0.410 \\
        $\vP^2=1$, $n=2$ & $1.462(20)$ & 0.188 [1] & 0.664 & $1.491(35)$ & 0.069 [1] & 0.793 \\
        $\vP^2=1$, $n=3$ & $1.498(10)$ & 0.002 [1] & 0.967 & $1.502(18)$ & 0.006 [1] & 0.937 \\
        \hline
        $\vP^2=2$, $n=0$ & $0.9694(46)$ & 0.120 [1] & 0.729 & $0.9751(92)$ & 0.232 [1] & 0.630 \\
        $\vP^2=2$, $n=1$ & $1.0153(39)$ & 0.491 [1] & 0.483 & $1.0197(74)$ & 0.731 [1] & 0.393 \\
        $\vP^2=2$, $n=2$ & $1.2746(66)$ & 0.006 [1] & 0.937 & $1.277(12)$ & 0.001 [1] & 0.977 \\
        $\vP^2=2$, $n=3$ & $1.3134(81)$ & 0.633 [1] & 0.426 & $1.319(14)$ & 0.509 [1] & 0.476 \\
        $\vP^2=2$, $n=4$ & $1.530(11)$ & 0.435 [1] & 0.510 & $1.541(19)$ & 0.367 [1] & 0.545 \\
        \hline
        $\vP^2=3$, $n=0$ & $1.0155(99)$ & 0.511 [1] & 0.475 & $1.024(18)$ & 0.377 [1] & 0.539 \\
        $\vP^2=3$, $n=1$ & $1.1083(62)$ & 0.055 [1] & 0.815 & $1.116(12)$ & 0.019 [1] & 0.891 \\
        $\vP^2=3$, $n=2$ & $1.596(17)$ & 0.050 [1] & 0.822 & $1.605(33)$ & 0.078 [1] & 0.781 \\
        \hline
        $\vP^2=4$, $n=0$ & $0.8367(94)$ & 0.007 [1] & 0.932 & $0.855(20)$ & 0.080 [1] & 0.777 \\
        $\vP^2=4$, $n=1$ & $1.055(15)$ & 0.923 [1] & 0.337 & $1.058(34)$ & 0.906 [1] & 0.341 \\
        $\vP^2=4$, $n=2$ & $1.1731(57)$ & 1.323 [1] & 0.250 & $1.181(11)$ & 1.079 [1] & 0.299 \\
        $\vP^2=4$, $n=3$ & $1.422(10)$ & 2.403 [1] & 0.121 & $1.424(20)$ & 2.397 [1] & 0.122 \\
        \hline
    \end{tabular}
    \caption{Continuum extrapolations of the finite-volume energy levels using both $\mathcal{O}(a^2)$ and $\mathcal{O}(a)$ ans\"atze at fixed $M_\pi L=6.40$.}
    \label{tab:cl-energies-comparison}
\end{table}

\FloatBarrier
\bibliographystyle{JHEP}
\bibliography{ref}

\end{document}